\documentclass[aps,twocolumn,floatfix]{revtex4-1}
\usepackage{amssymb}
\usepackage{graphicx}
\usepackage{caption}
\usepackage{subcaption}
\usepackage{multirow}
\usepackage[table,xcdraw]{xcolor}
\usepackage{soul}
\usepackage{kantlipsum}
\usepackage{amsmath}
\usepackage{amsfonts}
\usepackage{amssymb}
\usepackage{dcolumn} 
\usepackage{braket}
\usepackage{booktabs}
\usepackage{multirow}
\usepackage[table]{xcolor}
\usepackage{dsfont}
\usepackage{times}
\usepackage{epsfig}
\usepackage{xcolor}
\usepackage{tikz}
\usepackage{float}
\usepackage{color}
\usepackage{inputenc}
\usepackage{booktabs}
\usepackage{pgfplotstable}
\def\bx {{\bf x}}
\def\by {{\bf y}}

\def\bkappa {\boldsymbol{\kappa}}

\def\btau {\boldsymbol{\tau}}

\newcommand{\beq}{\begin{equation}}
\newcommand{\eeq}{\end{equation}}
\newcommand{\ba}{\begin{eqnarray}}
\newcommand{\ea}{\end{eqnarray}}

\input epsf

\usepackage[english]{babel}

\begin{document}
 \title[]{\hspace{1.5cm} Geometrically navigating topological plate modes \newline around
   gentle and sharp bends}
\author{Mehul P. Makwana$^{1, 2}$, Richard V. Craster$^{1}$}
\affiliation{$^1$ Department of Mathematics, Imperial College London, London SW7 2AZ, UK }
\affiliation{$^2$ Multiwave Technologies AG, 3 Chemin du Pr\^{e} Fleuri, 1228, Geneva, Switzerland}

\begin{abstract}

  Predictive theory to geometrically engineer devices and materials in
  continuum systems to have desired topological-like effects is
  developed here by bridging the gap between quantum and continuum
  mechanical descriptions. A structured elastic plate, a bosonic-like system
  in the language of quantum mechanics, is shown to exhibit
  topological valley modes despite the system having no direct
  physical connection to quantum effects.  We emphasise a predictive,
  first-principle, approach, the strength of which is demonstrated by
  the ability to design well-defined  broadband edge states, resistant to backscatter, using
  geometric differences; the mechanism underlying energy transfer around gentle and sharp
  corners is described.  Using perturbation methods and group theory, several
  distinct cases of symmetry-induced Dirac cones which when gapped
  yield non-trivial band-gaps are identified and classified. 
  The propagative behavior of the edge states around gentle or sharp bends  depends strongly upon the symmetry class of the bulk media and we illustrate this via numerical simulations. 
 
\end{abstract}
\maketitle

\section{Introduction}

There has been considerable recent activity in
 wave phenomena motivated through topological effects: The critical
 realisation has been that fundamental ideas
 originating in topological insulators and quantum mechanics
 \cite{Kane_Mele_2005,hasan10a, Ren_2016}, based around the Schr\"odinger equation, carry
 across, in some regards, to continuous wave systems based around,
 say, 
the Maxwell equations, such as topological photonic and phononic
 crystals \cite{lu14a, khanikaev_two-dimensional_2017}. Much of
 the recent continuum literature draws very heavily upon that from quantum mechanics and it is important to note that some care is required in
 this translation. 

Similar care is required in dealing with the
 delicacies and repercussions of group theoretic concepts
 \cite{dresselhaus08a,inui90a, Dai_2017, Lu_2014}.
   We will go back to first principles and
 elucidate key details in the translation process, and in group
 theory, highlighting and clarifying common
 issues that arise within structured elastic plates and by extension
 to other periodic crystal structures such as those in photonics or phononics. A
 recurrent theme throughout the article is the power of group theory
 in terms of clarifying and classifying, {\it a priori} without any explicit
 calculation, when certain effects will occur in classical waves.  

The fields of group theory and topology transcend specific physical
systems, hence the phenomena we describe translate widely. However, there are naturally
technical differences and we choose to illustrate our theory within
the context of flexural waves upon thin
 structured elastic plates \cite{mcphedran09a}, by doing so we emphasise the continuum nature of the
 model and show the generality of the basic ideas: It has no
 connection with quantum mechanics in either its formulation or theoretical basis. Thin plate
 flexural wave theories, as described in \cite{landau70a,graff75a} 
   are highly effective physical models for
 elastic waves in plates and have proved to be reliable in predictions of many wave phenomena for
 structured plates \cite{lefebvre17a}, plate models utilising Dirac
 cones in the style of graphene 
 \cite{torrent13a}, illustrating cloaking \cite{farhat09a}, negative refraction \cite{farhat10a} and valley edge states \cite{pal17a}. Typically flexural wave
 theory, for homogeneous plates, is quoted as only being accurate for wavelengths greater than $20$ times the
plate thickness \cite{rose04a} which is rather limiting. 
However, as shown in \cite{lefebvre17a}, it is the wavelength in the periodic system that actually matters, and that can be large compared to the plate thickness, even at high frequencies.
These plate models also act
 in practical terms 
 as motivation for seismic metamaterial applications
 \cite{brule14,colombi16a}. Gaining understanding of the 
  interaction of sub-wavelength arrays of resonators with an underlying
 plate \cite{colombi14a,williams15a} enabled these concepts to carry across to full
 vector elastic systems involving Rayleigh waves
 \cite{colombi16b}. Similarly, plate models are also
 critical in terms of highlighting features such as zero-frequency
 stopbands \cite{movchan07c} that can then be used to try and design
 broadband seismic
 phononic shields that can function at the long-wave and low-frequency
 regimes that are of importance in that context \cite{krodel15a,miniaci16a,achaoui17a}. 

The elastic plate model is ideal in terms of 
 describing and modelling 
topological effects as many results for point
scatterers are available explicitly \cite{evans07a}. A particularly pleasant
feature of analysis using this model is that the fundamental Green's function
 is, unlike in acoustics, electromagnetism and vector elasticity,
 non-singular and it remains bounded. This means we can
 concentrate cleanly upon issues such as group theory, and its influence upon
 the effects we see, without numerical distractions. The plate model, unlike
 acoustic and electromagnetic counterparts, has wave dispersion even
 when homogeneous and thus exemplifies the ubiquitous nature of the
 effects we discuss.

 There are two canonical types of topological insulators, those which
 preserve time reversal symmetry (TRS) and those which break it.  In
 quantum mechanics the design of a TRS insulator is contingent upon
 the fractional spin of the fermions. Quantitively, this condition
 requires that the time-reversal operator squares to $-1$; this is
 different to Newtonian systems that consist of spin-$1$
 bosonic-like particles. In the absence of spin-half particles for our
 system, we leverage the pseudospins inherent in hexagonal lattices
 that have broken inversion symmetry. The binary valley degree of
 freedom can be used to design valleytronic devices
 similar to those in spintronics by leveraging the valley-pseudospin in the manner of
 electron spin. The prohibition of backscattering is reliant upon
 there being no intervalley scattering hence the valley Hall effect \cite{dong_valley_2017, zhu_design_2018, xiao_valley-contrasting_2007, chen17a} of our system
 here is unlike the quantum Hall effect that breaks TRS; as our valley Hall insulators do not break TRS they are far more straightforward to realise.

In the topological context, care must be taken when
navigating waves around bends as we must prohibit hybridisation
of modes with opposite pseudo-spin. Transport of energy around corners in structured media
is of inherent interest \cite{mekis96a, ma_guiding_2015, chutinan_wider_2002}.  We simultaneously require
that the incoming edge mode can couple into a mode, that must exist,
along the interface after the corner. Critically the existence of that
mode is dependent upon the geometrical properties of the elementary
cells of the structured media and their relative arrangement.  
 Within these constraints we now proceed to construct 
 valley-dependent edge states that enable the suppression of intervalley
 scattering along the zigzag boundary, leading to valley-protected
 broadband robust transport around a bend.

We begin in Sec. \ref{sec:platonic} by explicitly recasting the continuum plate model into the language of quantum
mechanics, utilising a Hamiltonian description, whilst retaining elements
of the continuum language to bridge across the quantum and elastic plate 
communities. 
In Sec. \ref{sec:engineering} we demonstrate the geometrical differences in propagation around gentle and sharp bends which adjoin 
topologically distinct media.
 Concluding remarks are drawn together in section \ref{sec:conclusion}.

\section{Elastic plate crystals}
\label{sec:platonic}

We consider the elastic plate analogues of photonic, or
phononic, crystals where a homogeneous plate is given structure by a
lattice of defects which could be holes, pillars, mass-spring
resonators or elastic rods; familiar effects
such as forbidden frequency band-gaps and dynamic anisotropy all emerge within this plate system. We choose to use the simplest defects,
that is we either use clamped points, ``pins'', of zero radius or
mass-loading at a point and both are common idealisations for point
scatterers \cite{evans07a}. Resonators attached at the lattice
vertices 
\cite{pal17a,torrent13a,xiao12a} could easily be added into the
formalism we present, but introduce an unnecessary additional physical feature
associated with resonance; using point-scatterers, that are
either simple masses or constrained, demonstrate
 that resonance is not required in order to obtain topological
  effects. 

\subsection{Formulation}
The flexural wave modes that exist on an infinite elastic plate with
 constraints at lattice points 
are characterised by their vertical displacement, $\psi_{n
  \bkappa}({\bf x})$. The subscript notation denotes that
this field variable is dependent upon the Bloch-wavevector ${\bkappa}$
and $n$ is an index that numbers the eigenmodes. 

These displacement eigenmodes are governed by the
(non-dimensionalised) Kirchhoff-Love (K-L) equation
\begin{equation}
\left[H_{KL}-\mu(\bx)\omega_{\bkappa}^2\right]\psi_{n \bkappa}=F(\bx),
\label{eq:kirchoff}
\end{equation} 
where we introduce the operator $H_{KL}$ as 
\begin{equation}
H_{KL}=\nabla_{\bf x}^2\left[\beta\left(\bx \right)\nabla_{\bf x}^2 \right]
\label{eq:kirchhoff}
\end{equation}
 and the reaction forces at the point constraints $F(\bx)$ introduce
 the dependence upon the direct lattice. The quantities $\mu(\bx)$ and
 $\beta(\bx)$  represent variations in the non-dimensional mass per unit length and flexural rigidity of the
 plate respectively; their spatial dependence respects the same
 periodicity as the lattice of point constraints.


The simplest constraints are those of point mass-loading with 
the reaction forces proportional to the displacement 
 via an impedance coefficient and thus 
\beq
F(\bx)=\omega_{\bkappa}^2\sum_{\bf n}\sum_{p=1}^{P}
 M^{(p)}_{\bf n} \psi_{n \bkappa}(\bx)\delta\left({\bf x}-{\bf x}^{(p)}_{\bf n}\right).
\label{eq:ML_RHS}
\eeq 
Here ${\bf n}$ labels each elementary cell, containing $p=1...P$ constraints, that periodically repeats to create the infinite physical plate crystal. The mass in cell ${\bf n}$ at
point constraint $p$ is given by $M_{\bf n}^{(p)}$. This constraint
automatically encompasses the point pinned plate crystal, as the limit 
$\omega^2_{\bkappa}M^{(j)}_{\bf n}\rightarrow \infty$, where the
reaction forces are retained 
\beq 
F(\bx)=\sum_{\bf n}\sum_{p=1}^{P}F^{(p)}_{\bf n}\delta\left({\bf
    x}-{\bf x}^{(j)}_{\bf n}\right)
\label{eq:PP_RHS}\eeq 
but the displacement is constrained explicitly to be zero at the pins, i.e. $\psi_{n\bkappa}({\bf
  x}^{(p)}_{\bf n})=0$.


More generally, it is convenient to use the eigenstate notation for the K-L
equation as 
\beq
\hat{H} \ket{\psi_{n \bkappa}} = \hat{\mu}(\bx)\omega_{\bkappa}^2
\ket{\psi_{n \bkappa}} + F(\bx) \ket{\psi_{n \bkappa}}, 
\label{eq:simult_equations}\eeq 
where the component equation (\ref{eq:kirchoff}) is retrieved from
Eq. (\ref{eq:simult_equations}) using 
\beq
\hat{H} = \int \ket{\bx} H(\bx, \by) \bra{\by} d\bx d\by, \quad H(\bx, \by) = \delta(\bx - \by)H_{KL}.
\nonumber
\eeq 
In an infinite medium the displacements are Bloch eigenfunctions 
\beq
\psi_{n \bkappa}(\bx) = \braket{\bx|\psi_{n\bkappa}} = \exp\left(i\bkappa\cdot\bx \right) \braket{\bx|u_{n\bkappa}},
\nonumber
\eeq where $\ket{u_{n\bkappa}}$ is a periodic eigenstate.
The displacements satisfy the following completeness and orthogonality relations: 
\beq
\sum_{n \bkappa} \ket{\psi_{n\bkappa}} \bra{\psi_{n\bkappa}} = \hat{1}, \quad \braket{\psi_{n\bkappa}|\psi_{m\bkappa'}} = \delta_{mn} \delta_{\bkappa, \bkappa'}.
\label{eq:complete_orthogonal_set}
\eeq 
Due to the periodic arrangement of the inclusions, the displacement
response, in Eq. \eqref{eq:kirchoff},  naturally encourages a Fourier representation
\beq
\psi_{n\bkappa}(\bx)= \sum_{{\bf G}} W({\bf G}) \exp{\left(i({\bf
         G}-{\bkappa})\cdot {\bx} \right)}.
\label{eq:psi_FS}
\eeq 
as a sum over reciprocal lattice vectors ${\bf G}$. This gives the
formal solution in reciprocal space via 
\begin{multline}
\left( \vert {\bf G}-{\bkappa} \vert^4-\omega_{\bkappa}^2 \right) W({\bf G})
=\\
\frac{\omega_{\bkappa}}{A_{\text{PC}}}\sum_{p=1}^{P}M^{(p)}_{\bf I} \psi_{n\bkappa}\left(\bx^{(p)}_{\bf I} \right) \exp\left[-i({\bf G}-{\bkappa})\cdot {\bx}^{(p)}_{{\bf I}} \right],
\label{eq:transformed}
\end{multline} 
where ${\bf I}$ denotes an arbitrary reference cell in physical space, 
$A_{\text{PC}}$ is the area of the primitive cell
and, for clarity, we do not allow for spatial dependence of physical
parameters. This formulation is convenient for numerical simulation, and we use an
adaptation \cite{xiao12a} of the plane wave expansion method
\cite{johnson01a} to determine the eigenstates.

\subsection{Perturbation theory}
We now apply the ${\bf k}\cdot{\bf p}$ perturbation method \cite{Mei_2012, janssen_precise_2016} to the system \eqref{eq:kirchoff} whereby we retrieve the perturbed eigensolutions as a function of those at a reference point in Fourier space. Initially, we define a new complete orthogonal set, namely, the Kohn-Luttinger functions as,
\beq
\chi_{n\bkappa}(\bx) = \exp\left(i\bkappa\cdot\bx \right) \chi_{n\bkappa_0}(\bx),
\label{eq:K_L_functions} \eeq where $\bkappa_0$ is a fixed wavevector. Similar to $\psi_{n\bkappa}(\bx)$, they form a complete orthogonal basis set, \eqref{eq:complete_orthogonal_set}. Using the Kohn-Luttinger functions, we can expand any Bloch state, $\psi_{n\bkappa}(\bx)$, in the complete orthogonal basis set $\{ \chi_{j\bkappa}(\bx)\}$,
\begin{multline}
\ket{\psi_{n\bkappa}} = \sum_j A_{nj}(\bkappa) \ket{\chi_{j\bkappa}} =  \\ 
\exp\left(i\Delta\bkappa\cdot\bx \right) \sum_j A_{nj}(\bkappa) \ket{\psi_{j\bkappa_0}}
\label{eq:psi_expansion} \end{multline} and $\Delta \bkappa = \bkappa
- \bkappa_0$. After inserting the expansion into the governing
equation we obtain the integrand, 
\begin{multline}
\exp\left(i\Delta\bkappa\cdot\bx \right)\sum_j A_{nj}(\bkappa) [(\omega_{\bkappa_0}^2-\omega_{\bkappa}^2) \times \\
\left[\mu(\bx) + \sum_{{\bf N}, p}
 M^{(p)}_{\bf N} \delta\left({\bf x}-{\bf x}^{(p)}_{\bf N}\right) \right]
+ \\
\Delta\bkappa \cdot \hat{{\bf p}}_{\bx} 
+ \mathcal{O} \left(|\Delta\bkappa|^2 \right)]\psi_{j\bkappa_0}(\bx)=0.
\label{eq:expansion_gov}\end{multline} 
The momentum operator is explicitly given by 
\beq
\hat{{\bf p}}_{\bx} = 2i\left[\nabla^2_{\bx}\beta(\bx)\nabla_{\bx} + 3 \nabla_{\bx}\beta(\bx)\nabla_{\bx}^2 + 2\beta(\bx)\nabla^3_{\bx}\right]
\label{eq:alpha_defn} \eeq 
and the orthogonality of the eigensolutions, \eqref{eq:complete_orthogonal_set}, is written as,
\begin{multline}
\braket{\psi_{n\bkappa}|\psi_{m\bkappa'}} = 
\\
\frac{(2\pi)^2}{A_{PC}} \int_{\text{PC}}\psi^*_{n\bkappa}(\bx)\mu(\bx)\psi_{m\bkappa'}(\bx) d\bx = \delta_{mn} \delta_{\bkappa,\bkappa'},
\label{eq:orthogonality_integral} \end{multline} 
where the integral is taken over the primitive cell. Motivated by the orthogonality relation we multiply \eqref{eq:expansion_gov} by $\psi^*_{l\bkappa_0}(\bx)$ and integrate  over the primitive cell to get
\begin{multline}
\sum_j \left(H_{lj} - (\omega_{\bkappa}^2-\omega_{\bkappa_0}^2) \Lambda_{lj}\right)A_{nj}(\bkappa)=0, \\
\quad H_{lj} = \Delta\bkappa \cdot {\bf p}_{lj} + \mathcal{O}(|\Delta\bkappa|^2), \quad {\bf p}_{lj} = \braket{\psi_{l\bkappa_0}|\hat{\bf{p}}_{\bx}|\psi_{j\bkappa_0}}, \\
\Lambda_{lj}=\delta_{lj} + \sum_{p} M^{(p)}_{{\bf I}} \psi^*_{l\bkappa_0}\left(\bx^{(p)}_{{\bf I}}\right) \psi_{j\bkappa_0}\left(\bx^{(p)}_{{\bf I}}\right)
\label{eq:expanded_equation}\end{multline}  Nontrivial eigensolutions exist if the following secular equation is satisfied,
\beq
\det\left[H-(\omega_{\bkappa}^2-\omega_{\bkappa_0}^2)\Lambda \right] = 0.
\label{eq:disp_relation}\eeq To retrieve the full dispersion relation, the entire basis set at $\bkappa_0$, would need to be considered. However, if we are solely interested in linear dispersions around Dirac- and Dirac-like cones then the summation is limited to the degenerate Bloch states (as these eigensolutions, collectively, form an orthogonal degenerate subspace). Hence after neglecting second- and higher-order terms, \eqref{eq:expanded_equation} is simplified to,
\begin{multline}
\sum^d_{j=1} \left[H^{(1)}_{lj} - 2\omega_{\bkappa_0} \Delta \omega \Lambda_{lj} \right]A_{nj}(\bkappa)=0, \\
\Delta \omega = \omega_{\bkappa}-\omega_{\bkappa_0} = \Delta\bkappa \cdot \nabla_{\bkappa} \omega_{\bkappa_0} + \mathcal{O}(|\Delta\bkappa|^2), \\
\quad H^{(1)}_{lj} = \Delta\bkappa \cdot{\bf p}_{lj}, 
\label{eq:reduced_linear_equation}\end{multline} where $d$ is the degree of degeneracy. From this we see that the first-order term is exclusively determined by the strength of the coupling between the degenerate Bloch states. Other non-degenerate states contribute to higher-order corrections \cite{dresselhaus08a}.
No assumption has been made on the origins of the degeneracy; therefore the above equation is applicable to both, essential and accidental, degeneracies with linear dispersive behaviour. As an alternative to the Kohn-Luttinger model we could have opted to expand $\ket{\psi_{n\bkappa_0}}$ into its constituent plane-waves \cite{ochiai_photonic_2012} to obtain the same result. 

\subsection{Symmetry induced Dirac cones}
In this subsection we rectify and extend basic group theoretical concepts, found in \cite{lu_dirac_2014}, that aid us in the reduction of the Hamiltonian. All the space group elements, which leave our lattice invariant, are succinctly written as $\left\{R, \btau \right\}$; where $R$ denotes a point group symmetry element and $\btau$ corresponds to a lattice translation. The triangular-like periodic structures, under consideration here, belong to a symmorphic space group $G$; hence all $\left\{R, \btau \right\} \in G$ are compound symmetries obtained by combining a point group element and a primitive lattice translation. Specifically, the elements in $G$ are separable, $\left\{R, \btau \right\}=\left\{R, 0 \right\}\left\{\varepsilon, \btau \right\}$; where $\btau$ denotes a Bravais lattice translation. It follows that the plate wavefunctions satisfy the following equalities,
\begin{multline}
\hat{P}_{\left\{R, \btau \right\}}\psi_{n\bkappa}({\bf x}) = \hat{P}_{\left\{R, 0\right\}}\hat{P}_{\left\{\varepsilon, \btau \right\}} \psi_{n\bkappa}({\bf x})
=  \\
\exp\left(i\hat{R}\bkappa\cdot\btau \right)\psi_{n \hat{R}\bkappa}({\bf x}),
\label{eq:symm_wavefunction}
\end{multline}  where $\hat{R}$ denotes the operator form of $R$. If
we are at a high-symmetry point in the Brillouin zone then $\hat{R}
\bkappa = \bkappa \mod {\bf G}$, for many different $\hat{R}$,  where
${\bf G}$ is an arbitrary reciprocal lattice vector. Each $\hat{R}$,
which satisfies this transformation property at $\bkappa$, belongs to
the group of the wavevector, denoted by $G_{\bkappa}$. The wavevector
group of highest symmetry is $G_{\Gamma}$ which is isomorphic to the
factor group $G/T$; $T$ is the translation subgroup. Hence, any
non-zero wavevector group is a (normal) subgroup of $G_{\Gamma}$; notationally this is written as $G_{\bkappa} \vartriangleleft G_{\Gamma}$ . At
high-symmetry points in Fourier-space, deterministic degeneracies
form, which yield a degenerate set of eigenfunctions $\left\{\psi_{n
    \hat{R}\bkappa}({\bf x}) \right\}$, which correspond to the same
frequency. In the following two subsections we shall use the point group symmetries, at the $KK^\prime$ valleys, to simplify the Hamiltonian found in Eq. \eqref{eq:reduced_linear_equation}.

\begin{figure} [h!]
	\centering
\begin{tabular}{ll}
(a) Physical space & \qquad\qquad (b) Reciprocal space
\end{tabular}
\begin{tabular}{ll}
		\includegraphics[width=5cm]{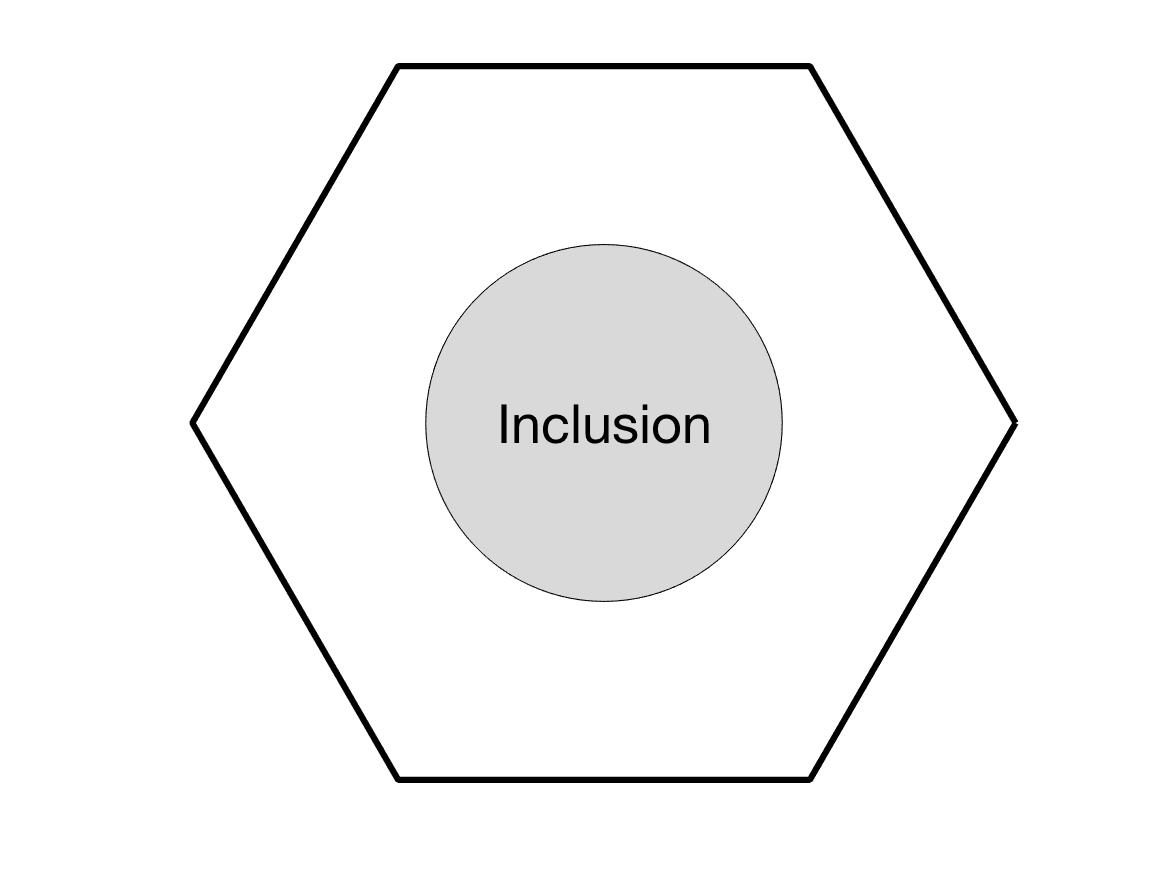} &
	\hspace{-0.75cm}	\includegraphics[width=5cm]{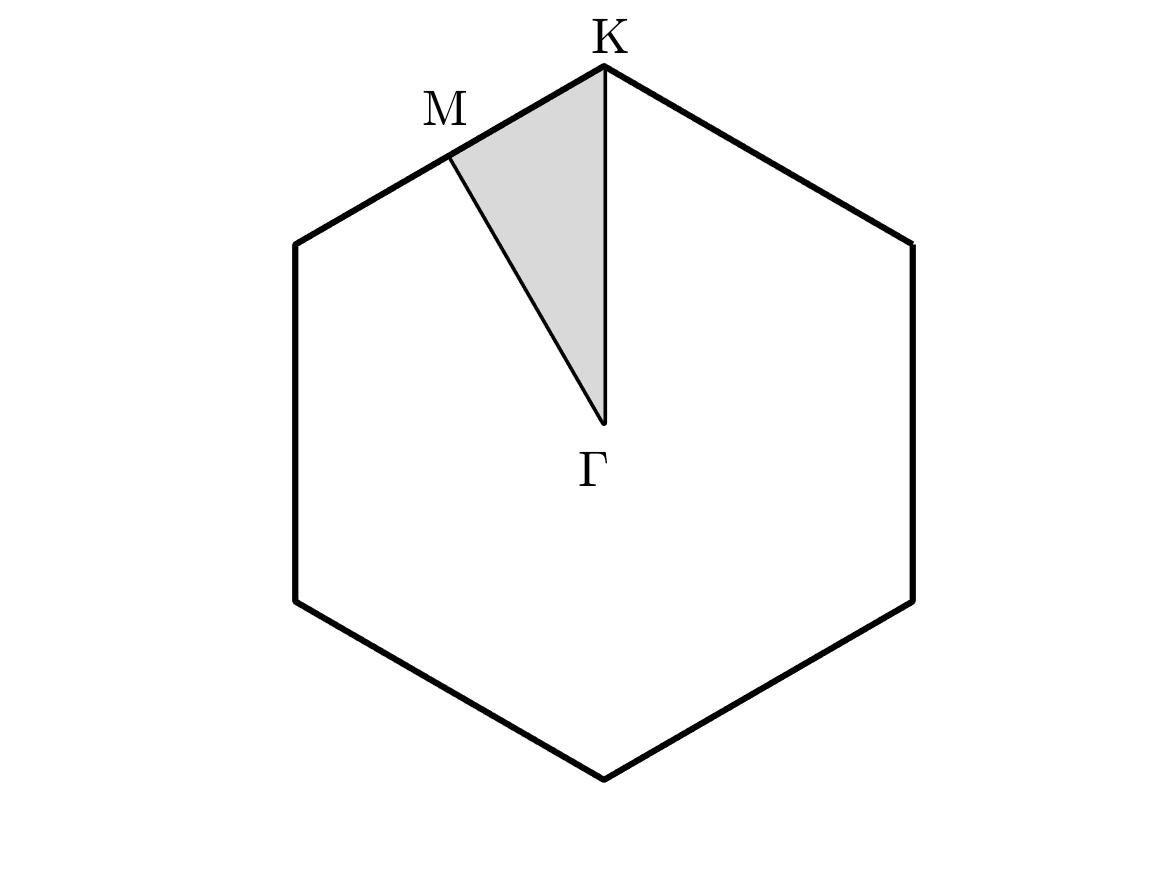}
\end{tabular}
	\caption{Wigner-Seitz cell in physical and reciprocal
          space. Shaded region, in the latter, indicates the IBZ.} 
\label{fig:wigner}
\end{figure}

\begin{figure} [h!]
	\centering
\begin{tabular}{ll}
(a) Physical space & \qquad\qquad (b) Reciprocal space
\end{tabular}
\begin{tabular}{ll}
		\includegraphics[width=4.5cm]{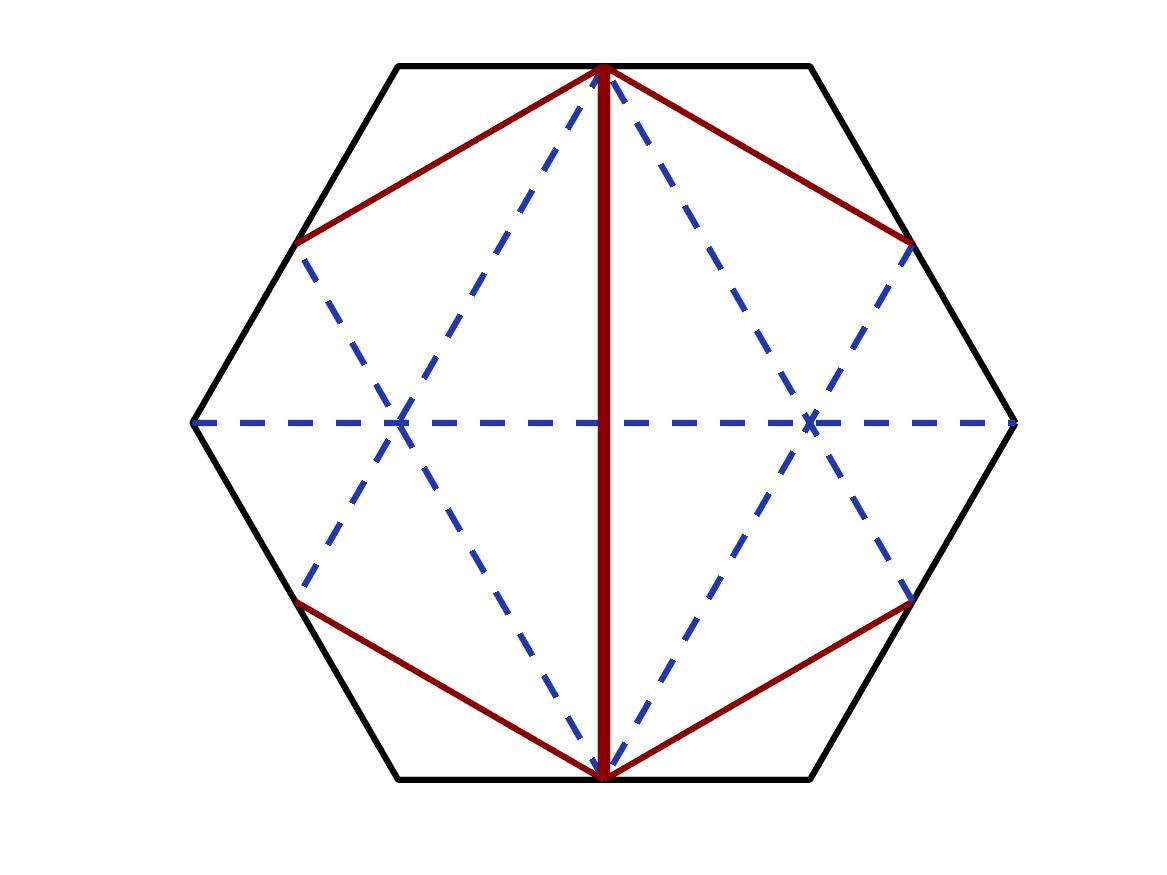} &
	\hspace{-0.75cm}	\includegraphics[width=4.5cm]{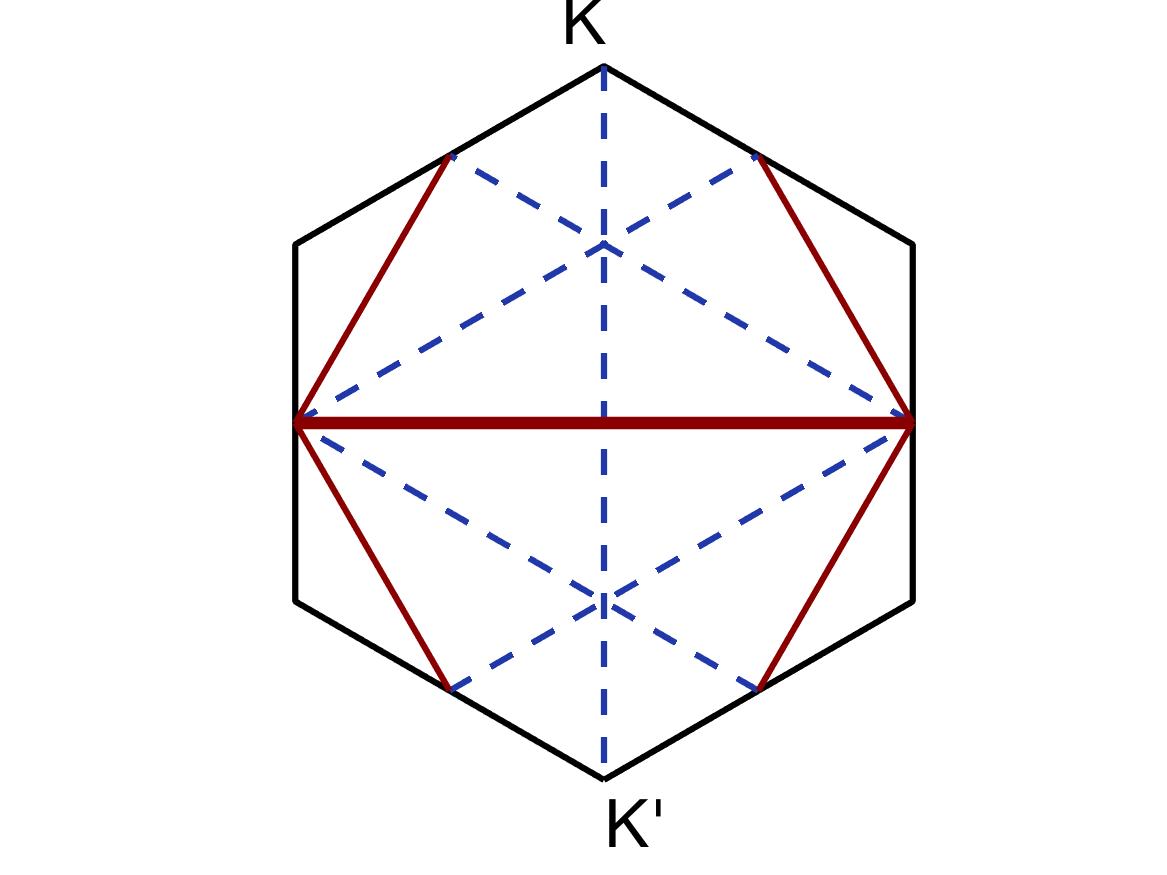}
\end{tabular}
	\caption{The two sets of reflectional symmetries:
          $\{\sigma_{d}\} $ and $\{\sigma_v\}$ (solid lines) in the
        physical and reciprocal spaces respectively. Similarly 
          $\{\sigma_{v}\},\{\sigma_{d}\}$ are shown dashed in physical and
          reciprocal space \cite{vasseur13a}. }
\label{fig:reflectional}
\end{figure}

Symmetries for the Brillouin zone of an unperturbed system are given by the group $C_{6v} = C_6 + 3\sigma_v (+ 3\sigma_d)$, where the two distinct families of mirror symmetries, namely the vertical ($\sigma_v$) and dihedral ($\sigma_d$), are shown in Fig. \ref{fig:reflectional}.
The corners of the Brillouin zone are associated with the point
groups $C_{3v} = C_3 + 3\sigma_v,~ C_3$; where the dependency on
$3\sigma_d$ (if present) is dropped as the symmetry does not
contribute to gapping the Dirac cone. For the point group
$C_{3v}$ (see table \ref{C3v}) we ascertain, immediately, that the
$KK^\prime$ valleys always support Dirac cones. The symmetries at $\Gamma$
are also immediately found, as $C_{3v} \vartriangleleft G_{\Gamma}$, therefore $G_{\Gamma} = C_{6v}$ or $C_{3v}$. Examples, of hexagonal lattices that have these symmetries are shown, as the first two rows, in table \ref{tab:hexagonal_DC}. For continuous systems, containing a periodic array of scatterers, it is sufficient to solely consider a hexagonal cell with a single inclusion; the benefits of the honeycomb or Kagome structures are in the additional degrees of freedom that they offer when analysing discrete analogues of continuous systems \cite{Zhang_Niu_2017, Liu_Xu_Zhang_2017, Liu_Xu_2017, Guo_2009, ni_topological_2017}.
Hence, for simplicity, we primarily consider the single `inclusion' case from hereon. 
Additionally, to have explicit solutions we use point scatterers in place of finite radii inclusions; where the point scatterers are placed at the vertices of the `inclusion'. It is also sufficient to consider the illustrative examples, of triangular and hexagonal inclusions, as all permutations that yield Dirac cones (and which gap them) are distilled down into these two distinct shapes. 

Unlike the point group $C_{3v}$, the $C_3$ point group supports a deterministic Dirac cone if and only if $G_{\Gamma} = C_6$. In this instance, the two complex one-dimensional representations in table \ref{C3} stick together due to the presence of TRS. 
Let us demonstrate this fact by considering the point group of an arbitrary element $\bkappa_0$. If an element $g \in G_{\Gamma}$ such that $g:\bkappa_0 \rightarrow -\bkappa_0$ then clearly $g^2 \in G_{\bkappa_0}$. Now, if the inversion operator $\hat{\pi} \in G_{\bkappa_0}$ then $\{g\} \in G_{\bkappa_0}$ and, trivially, we are at $\Gamma$ where we obtain a non-degenerate quadratic curve. However if $\hat{\pi} \notin G_{\bkappa_0}$ then $g$ may or may not belong to $G_{\bkappa_0}$, hence a test is required to discern which of the above cases is satisfied. The test conducted is known as Herring's criterion \cite{inui90a} and is succinctly written as,
\beq
\sum_{\{g\}} \chi \left(g^2 \right) = \begin{cases} h_g, & \text{no additional degeneracy}, \\
 0, -h_g & \text{doubling of degeneracy},
\label{eq:Herring}\end{cases}
\eeq 
where $\chi \left(g^2 \right)$ is a character associated to a specific irreducible representation (IR) in $G_{\bkappa_0}$ and $h_g$ is the number of elements in $\{ g\}$. This test is used to determine whether or not TRS introduces any additional degeneracies. Returning to our specific case, let $\bkappa_0 = K$ or $K^\prime$; now since  $\hat{\pi} \notin C_3$ we proceed with Herring's criterion. If $G_{\Gamma} = C_6$ then $\{ g\} = \{C_2, C_6, C^{-1}_6 \}$. In other words, these are the only elements within $C_6$ which transform $K$ to its TR counterpart $K^\prime$ and vice-versa. Squaring these elements we find that $\{C_2^2, C_6^2, C^{-2}_6\} = \{E, C_3, C^2_3\}$; next we perform Herring's criterion by using table \ref{C3},
\beq
\chi \left(E \right) + \chi \left(C_3 \right) + \chi \left(C^2_3 \right) = 1+\epsilon + \epsilon^2 = 0,
\eeq where $\epsilon = \exp\left(2\pi i/3 \right)$; this is the second case in \eqref{eq:Herring} therefore if $G_{\Gamma} = C_6$ we get a double degeneracy due to the presence of TRS. Similarly, this test can be carried out for  $G_{\Gamma} = C_{3v}$, where $\{ g\} = \{3 \sigma_v \}$, $\sum \chi \left(g^2 \right) = h_g$ which is the first case hence when $\{G_{\Gamma}, G_{KK^\prime}\} = \{C_{3v}, C_{3}\}$ we obtain a non-degenerate quadratic curve. 

In summary, the three sets of symmetries that yield deterministic Dirac cones are $\{G_{\Gamma}, G_{KK^\prime}\} = \{C_{6v}, C_{3v}\}$, $\{C_{3v}, C_{3v}\}$, $\{C_{6}, C_{3}\}$. Examples of structures yielding these symmetry sets are shown in table \ref{tab:hexagonal_DC}. 

\begin{table}[h!]
\centering
\caption{Hexagonal Lattice Dirac Cones}
\label{my-label}
\begin{tabular}{|c|c|c|c|c|}
\hline
\rowcolor[HTML]{EFEFEF} 
Case & $G_{\Gamma}$ & $G_{K, K^\prime}$ & Example Inclusions   & $K \leftrightarrow K^\prime$    \\ \hline
(i) & $C_{6v}$     & $C_{3v}$   & Hexagonal, $\lambda = \epsilon^n$   & $\{\sigma_v\}, \hat{\pi}$                       \\
(ii) &$C_{3v}$     & $C_{3v}$   & Triangular, $\lambda = \epsilon^{2n}$ & $\{\sigma_v\}$ \\
(iii) & $C_{6}$      & $C_{3}$    & Hexagonal, $\lambda \neq \epsilon^n$ & $\hat{\pi}$ \\ \hline
\end{tabular}
\label{tab:hexagonal_DC}
\caption*{The three cases: Here $\lambda$ is the angle of rotation away from the vertical axis and $\epsilon = \exp(i\pi/6)$.
The column $K \leftrightarrow K^\prime$ indicates whether the mirror $\{\sigma_v\}$ and/or inversion $\hat{\pi}$ symmetries map $K$ to $K^\prime$ (and vice versa). }
\end{table}


\subsection{Hamiltonian reduction}
By using the group theoretic principles espoused in the previous section we can further reduce the Hamiltonian of Eq. \eqref{eq:reduced_linear_equation}. Due to the invariance of a scalar product to symmetry operations, applying $\hat{R}$ to an arbitrary wavevector, $\bkappa$, is equivalent to applying its inverse operator $\hat{R}^{-1}$ to the physical space vector, $\bx$. Therefore for an $R \in G_{\bkappa_0}$, Eq. \eqref{eq:symm_wavefunction} can be written as, 
\begin{multline}
\hat{P}_{R} \psi_{j \bkappa_0} (\bx) = \psi_{j \hat{R}\bkappa_0} (\bx) = \psi_{j \bkappa_0} \left(\hat{R}^{-1}\bx \right) \\
= \sum_{i=1}^d D(R)_{i, j} \psi_{i \bkappa_0}(\bx),
\label{eq:symm_operator_affect} \end{multline} 
where $D(R)$ is an irreducible matrix representation (IMR) which relates the basis functions in an IR
\beq
\hat{P}_{R}\ket{\psi_{j \bkappa}}= \sum_{i=1}^d D(R)_{i, j} \ket{\psi_{i \bkappa}}. 
\label{eq:basis_function_definition}\eeq 
Motivated by Eq. \eqref{eq:symm_operator_affect}, we apply a change of variables, namely ${\by} = \hat{R}^{-1} {\bx}$, to ${\bf p}_{lj}$ in \eqref{eq:expanded_equation},
\beq
\begin{split}
{\bf p}_{lj} &= \int_{\text{PC}} \psi^*_{l \bkappa_0}\left(\by \right) \hat{{\bf p}}_{\by} \psi_{j \bkappa_0}\left(\by \right) d\by \\
 &= \hat{R}^{-1} \sum_{m, n} D^*(R)_{l, m} D(R)_{n, j} {\bf p}_{mn},
\end{split}\label{eq:matrix_elts_relationship}
\eeq 
as well as to $\Lambda_{lj}$,
\beq
\Lambda_{lj}= \sum_{m, n} D^*(R)_{l, m} D(R)_{n, j} \Lambda_{mn}.
\label{eq:Lambda_matrix_elts_relationship}
\eeq
Recall that, as we are dealing with orthogonal transformations, the Laplacian is invariant under a change of basis; hence \eqref{eq:matrix_elts_relationship} is true for any system with an even-ordered governing equation.  Eq. \eqref{eq:matrix_elts_relationship} relates different matrix elements within the first-order correction, hence the perturbed Hamiltonian is reduced using the point group symmetries at ${\bkappa}_0$. 

Directly from the form of ${\bf p}_{lj}$, in  \eqref{eq:expanded_equation}, we  immediately ascertain that the matrix ${\bf p}$ is Hermitian. Therefore, for the doubly degenerate case, of the Dirac cone we get ${\bf p}_{12} = {\bf p}^{*}_{21}$. Restricting ourselves to symmetry induced Dirac cones we need only to consider the point groups $C_3 = \{e, C_3, C^2_3\}$ and $C_{3v} = C_3 + 3\sigma_v$. The point group $C_3$ supports two independent complex one-dimensional IR's; see table \ref{C3}. As mentioned earlier, by Herring's criterion; the pair of one-dimensional complex IRs join to yield a double-degeneracy if (and only if) the point group at $\Gamma$ is $C_6$; the resulting two-dimensional IMRs of the rotation operators are identical, up to a similarity transformation, to those found in $C_{3v}$ (see App. \ref{sec:IMR}). As $G_{KK^\prime} = C_3$ or $C_{3v}$ and $C_3 \vartriangleleft C_{3v}$, we opt to initially reduce the Hamiltonian by using the rotation operators belonging to $C_3$; additionally, due to the equivalence of the representations $E$, for both $C_3$ and $C_{3v}$, we use the same IMR, see Eq. \eqref{eq:IR_E_C3_C3V} in App. \ref{sec:IMR}.

After substituting $\hat{R} = \pm \hat{C}_3$ into \eqref{eq:matrix_elts_relationship}, 
 we arrive at the following relations,
\beq
\begin{split}
{\bf p}_{11} = -{\bf p}_{22}, \quad {\bf p}_{12} = {\bf p}_{21}, \quad {\bf p}_{12} = -i \hat{\sigma}_y {\bf p}_{11}, \\
\Lambda_{11} = \Lambda_{22}, \quad \Lambda_{12}=\Lambda_{21}=0,
\label{eq:red_matrix_elts_C3}\end{split} \eeq where $\hat{\sigma}_y$ is the second Pauli matrix; the final equation in \eqref{eq:red_matrix_elts_C3} indicates that ${\bf p}_{11}$ and ${\bf p}_{12}$ are orthogonal. The Pauli matrices along with the identity matrix form a basis for the vector space of $2 \times 2$ Hermitian matrices; they are defined as,
\beq
\hat{\sigma}_a = \left[ \begin{array}{rr}
\delta_{a, z} & \delta_{a, x} - i \delta_{a, y} \\
\delta_{a, x} + i \delta_{a, y} & - \delta_{a, z} \end{array} \right],
\label{eq:Pauli_matrices}\eeq
where $a = x, y, z$. Using the relations \eqref{eq:red_matrix_elts_C3} we simplify the Hamiltonian, found in Eq. \eqref{eq:reduced_linear_equation}, to the following,
\beq
H^{(1)} = -|\Delta \bkappa| |{\bf p}_{11}| \begin{bmatrix} \begin{array}{rr}
-\cos \theta & \sin \theta  \\
\sin \theta & \cos \theta \end{array} \end{bmatrix},
\label{eq:red_Hamiltonian_C3} \eeq where $\theta$ is the angle between $\Delta \bkappa$ and ${\bf p}_{11}$. Somewhat unexpectedly  \eqref{eq:red_matrix_elts_C3} and \eqref{eq:red_Hamiltonian_C3} are identical to those for photonic crystals in \cite{lu_dirac_2014}. Herein we have demonstrated that they hold for all even-ordered two-dimensional systems. Using the reduced Hamiltonian and \eqref{eq:reduced_linear_equation} we find the gradient near the $KK^\prime$ valleys, in addition to, the first-order frequency correction,
\begin{multline}
2\omega_{{\bkappa}_0} \Delta \omega\Lambda_{11} = \pm |\Delta\bkappa| |{\bf p}_{11}|
\implies \frac{\Delta \omega}{|\Delta\bkappa|} = \pm\frac{|{\bf p}_{11}|}{2\omega_{{\bkappa}_0}}\Lambda_{11}.
\label{eq:linear_slope_K} \end{multline}  
Note Eq. \eqref{eq:linear_slope_K} is angular-independent; therefore the linear slopes, near Dirac cones, are independent of the reflectional properties of the system. However, if we are dealing with a system which has $C_{3v}$ symmetry, at the $KK^\prime$ points, then the set of three equivalent reflections $3\sigma_v$ are used to simplify the Hamiltonian \eqref{eq:red_Hamiltonian_C3} further by specifying the angular component.  

The IMR associated to the element $\sigma_v \in C_{3v}$ is proportional to the third Pauli matrix $\hat{\sigma_z}$ (see App. \ref{sec:IMR} for more details). Therefore, if $G_{KK^\prime} = C_{3v}$ we substitute $\hat{R} = \hat{\sigma}_z$ into Eq.  \eqref{eq:matrix_elts_relationship} which yields,
\beq
H^{(1)}_{\text{unpert}} = \tau_z v_D(\hat{\sigma}_z \Delta\kappa_x - \hat{\sigma}_x \Delta\kappa_y), 
\label{eq:Dirac_Hamiltonian}\eeq where the group velocity is,
\beq
v_D = \frac{|{\bf p}_{11}|}{2\omega_{{\bkappa}_0}}\Lambda_{11},
\eeq  
$\tau_z = \pm 1$ is the valley spin index, which corresponds to the $KK^\prime$ valleys, respectively. The connection between the $\tau_z$ term and the $KK^\prime$ valleys is seen explicitly by substituting in the binary rotation about the $z$-axis $C_2$ into \eqref{eq:matrix_elts_relationship}; this transforms $K$ into $K^\prime$. The plate Hamiltonian \eqref{eq:Dirac_Hamiltonian} resembles that of a massless Dirac fermion; it acts upon the amplitudes of the Kohn-Luttinger functions  \eqref{eq:K_L_functions} which, themselves, have been expanded around a degenerate pair of eigenstates \eqref{eq:psi_expansion}. Note that, alternatively, the Hamiltonian could have been reduced using selection rules \cite{mei_first-principles_2012, dresselhaus08a}.

Due to the simplified form of the reduced Hamiltonian, \eqref{eq:red_Hamiltonian_C3}, the Berry phase is immediately retrievable \cite{Berry_1983, Xiao_2010}. This phase factor, defined as, 
 \begin{multline}
\phi = i\oint \braket{\psi_{n \bkappa}| \nabla_{\bx}|\psi_{n \bkappa}} 
\cdot d{\bf l} = \\
i\oint \sum_j A^*_{nj}(\bkappa) \nabla_{\bkappa} A_{nj}(\bkappa)\cdot d{\bf l}, \quad 
\label{eq:Berry_phase} \end{multline} 
impacts wave transport properties when there is interaction between different Bloch eigenstates. The Kohn-Luttinger functions, in \eqref{eq:Berry_phase}, are analytically derivable by substituting the perturbed Hamiltonian \eqref{eq:red_Hamiltonian_C3} into the eigenvalue problem \eqref{eq:reduced_linear_equation}. In turn, these are substituted into the Berry phase formula \eqref{eq:Berry_phase} thereby yielding the expected result, for Dirac cones, $\phi = \pi$. The relevance of this result was shown in \cite{mei_first-principles_2012} where for Dirac-like cones $\phi = 0$ whilst for pure Dirac cones, $\phi =\pm \pi$. In a perfectly periodic medium, both, Dirac and Dirac-like cones are associated to perfect transmission of an incident wave through the medium; this is due to the locally linear curves which yield dispersionless waves. The physical ramifications of the difference in Berry phase present themselves in the presence of a defect whereby a zero Berry phase is associated with normal localisation whilst a system with a Berry phase of $\pm\pi$ implies antilocalisation effects. Under PT-symmetry, the Berry phase is quantized such that $\phi = \pi n$ hence no perturbation can continuously change it. In order to destroy a Dirac point, you either, move two Dirac points together which have opposite flux or you break parity and/or TRS. In the next section we shall break parity symmetry to gap the Dirac point, thereby yielding non-trivial band gaps.

\begin{figure}[h!]
\hspace{-0.5cm}
\includegraphics[width=9cm]{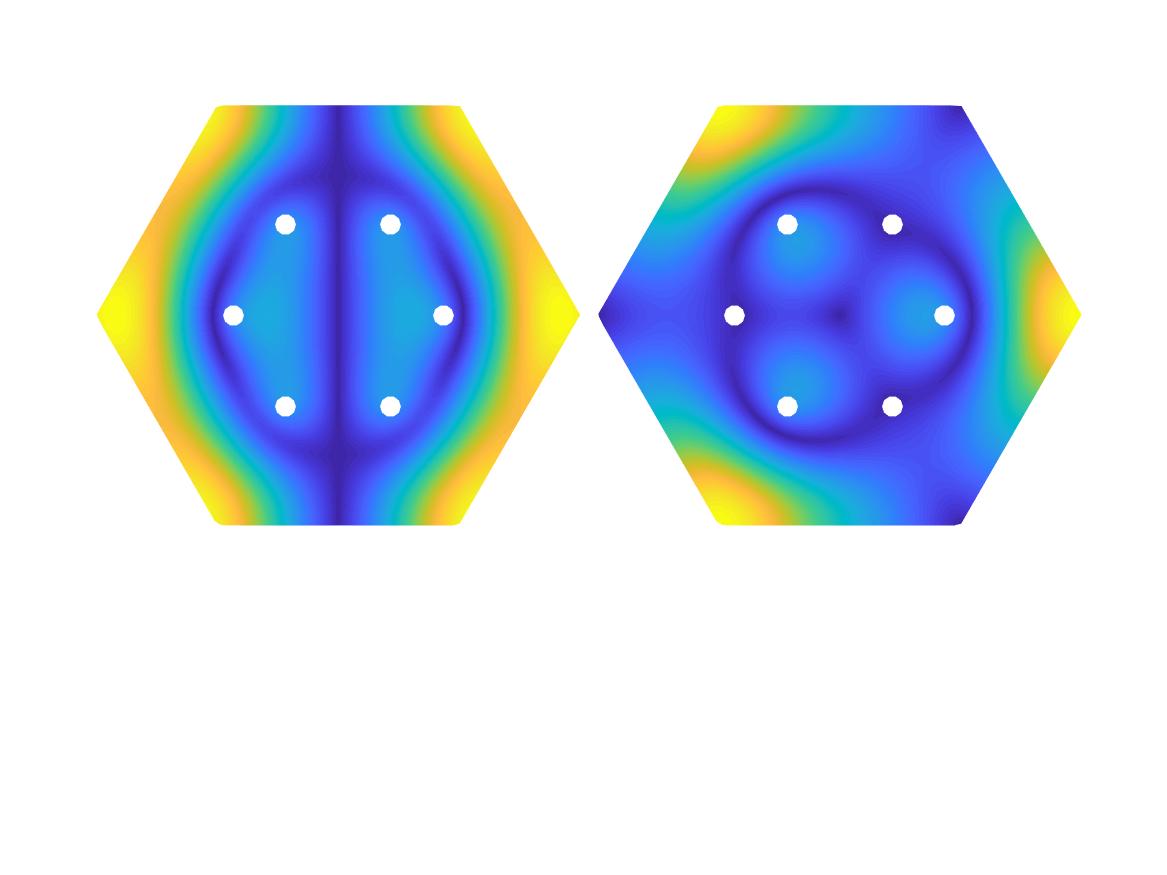}
\vspace{-2cm}
\caption{Symmetry breaking at $K$: the left and right panels show
          the absolute value of the displacement at the ungapped, and
          gapped Dirac point of case (i), shown in Fig. \ref{fig:C6v}, for
          $\omega=16.5$. }
\label{fig:wavefield}
\end{figure}

\begin{figure}[h!]
\hspace{-0.5cm}
\includegraphics[width=9cm]{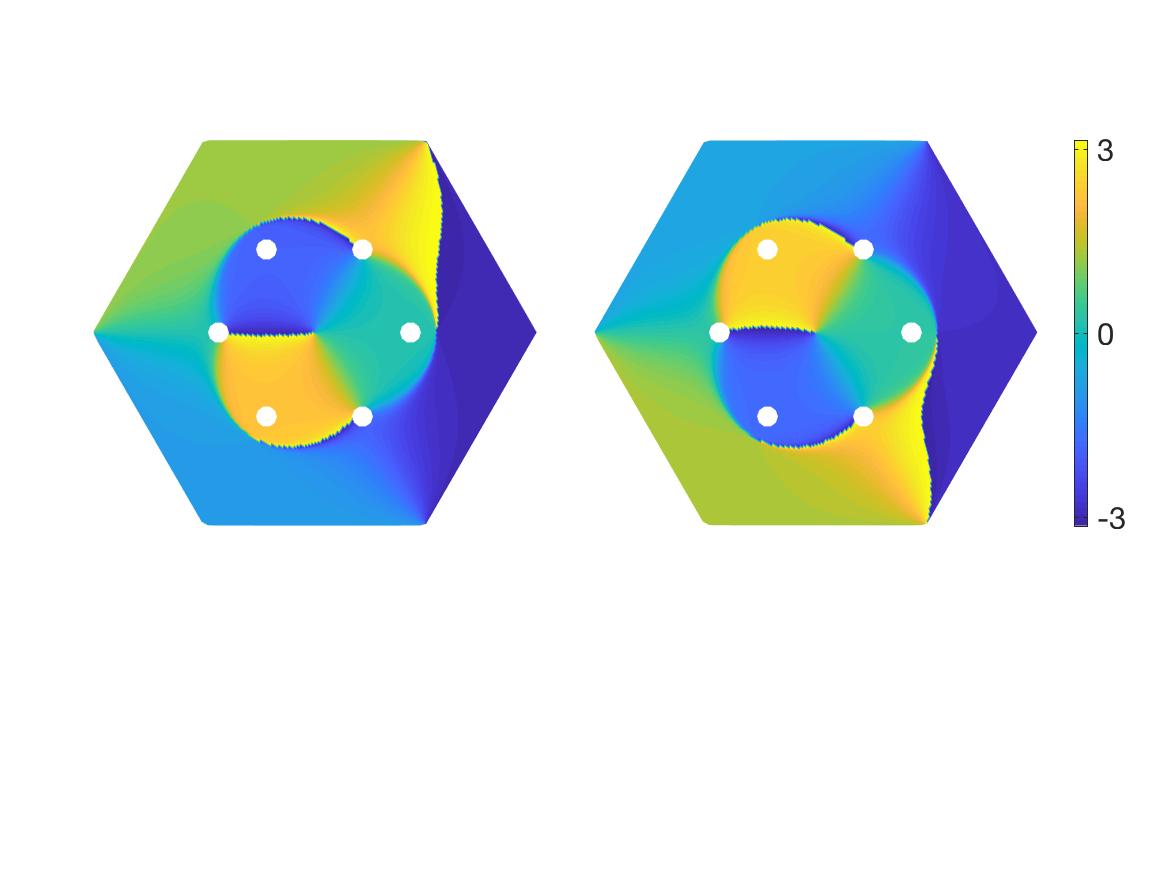}
\vspace{-2cm}
\caption{The chirality of the separated valley states at $K$ (left panel) and
  $K^\prime$ (right panel) for the gapped Dirac point of case (i) shown in
  Fig. \ref{fig:C6v}, $\omega=16.5$ showing the intrinsic
  circular-polarized orbital angular momentum using the phase
  distribution of the field.}
\label{fig:chirality}
\end{figure}

\vspace{0.5cm}
\subsection{Perturbed Hamiltonian and the valley Chern number}

We now demonstrate, from first-principles, how the Hamiltonian is altered when a Dirac point is gapped. We then immediately retrieve (local) topological quantities in the vicinity of the $KK^\prime$ valleys that indicate the existence of topological valley modes. 

\begin{figure} [h!]
\begin{tabular}{ll}
	\qquad (a) $\{C_{6v}, C_{3v} \}$ & \qquad \qquad \quad (b) $\{C_{3}, C_{3} \}$
	\\
	 Space Group: $p6m$ & \qquad \qquad  Space Group: $p3m1$
	
\end{tabular}

\begin{tabular}{ll}
		\includegraphics[width=3.25cm]{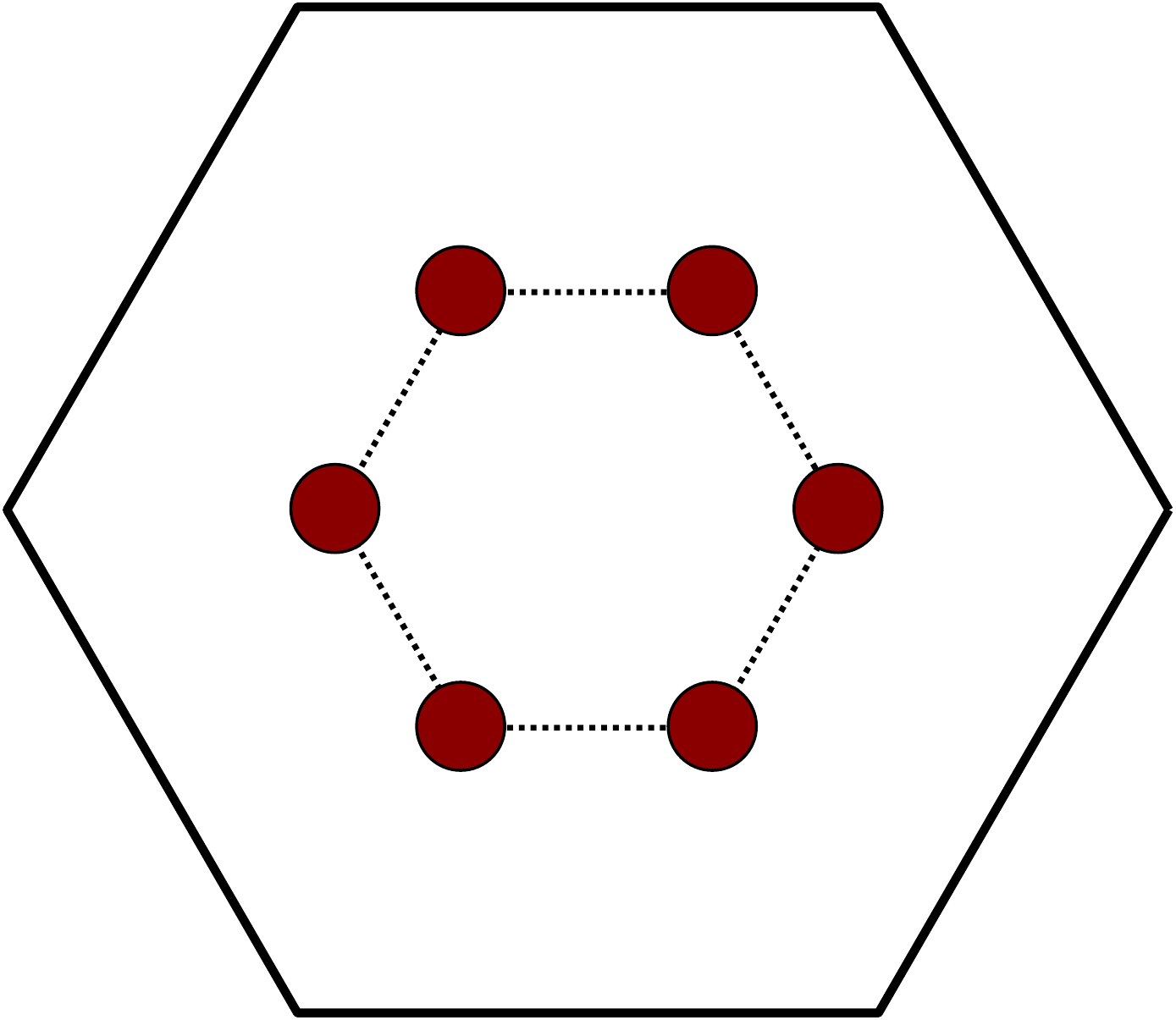} & 
	\hspace{2em} 
		\includegraphics[width=3.25cm]{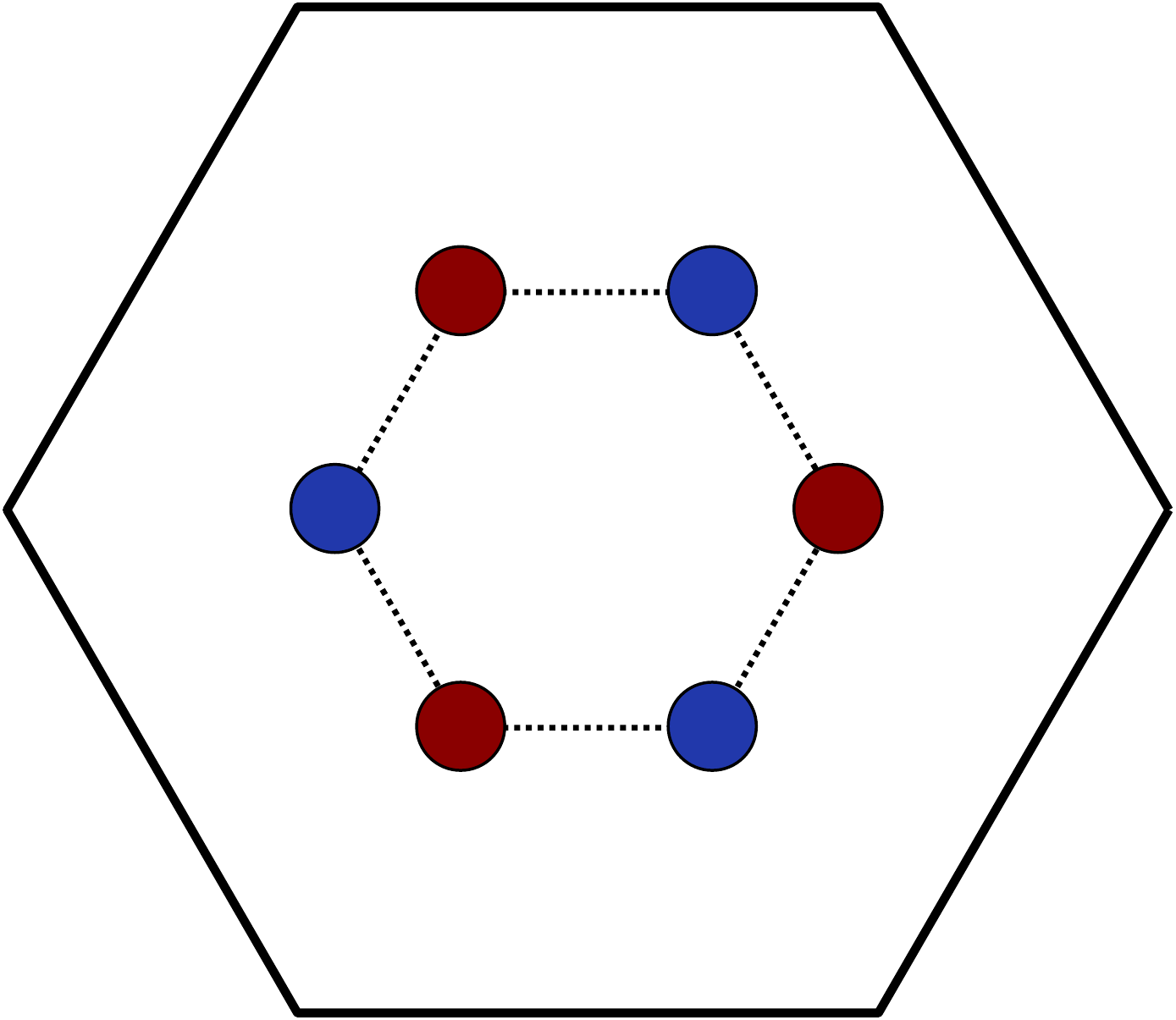} 
	\end{tabular}
		\includegraphics[width=8cm]{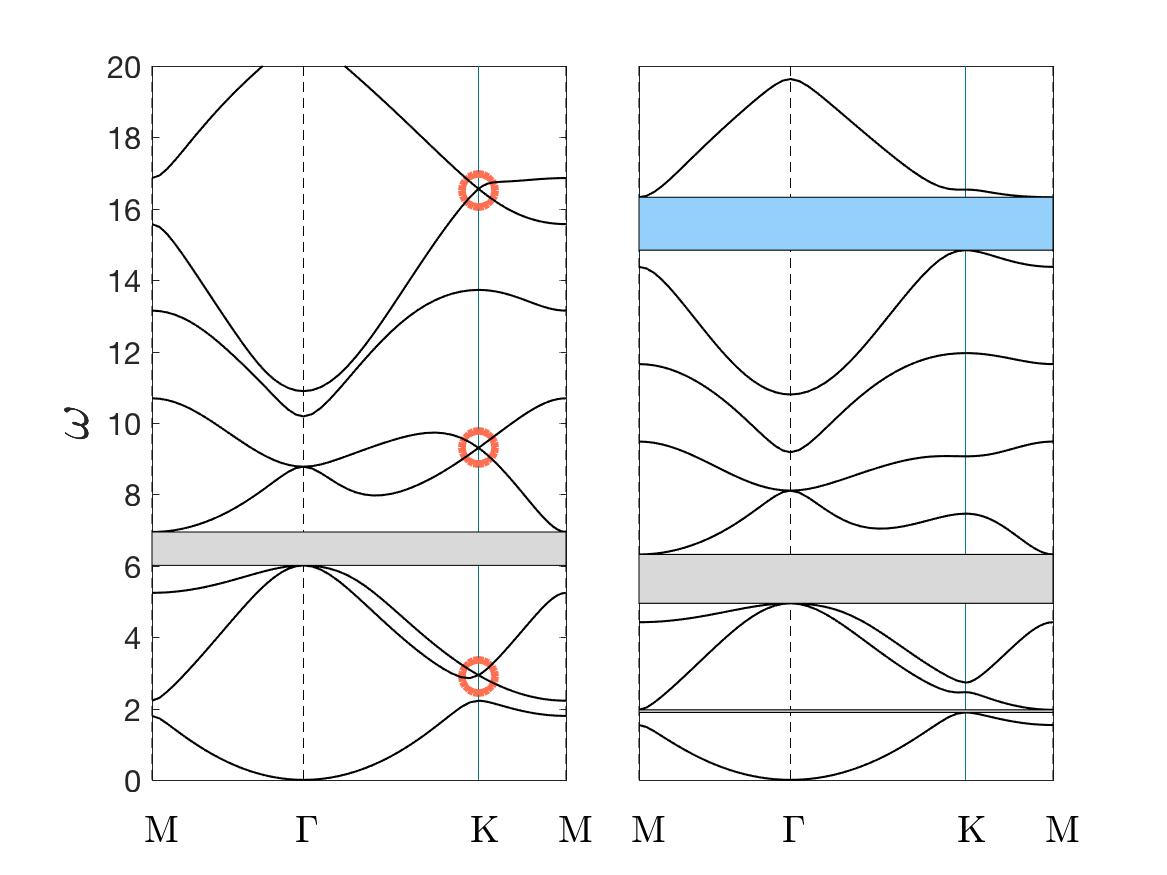}
	\caption{Case (i) in table
          \ref{tab:hexagonal_DC} generated by perturbing masses, showing
          the geometry, space groups and dispersion curves of (a)
          unperturbed and (b) perturbed cases: Unperturbed (perturbed)
          mass value $1$ ($2$) the norm of basis vectors is $2$ and
          distance from center of cell to masses is $0.5$ In (a) the
          Dirac points are circled and in (b) the new bandgap created
          by gapping the Dirac point is shaded blue.
} 
\label{fig:C6v}
\end{figure}


\begin{figure} [h!]
\begin{tabular}{ll}
	\qquad (a) $\{C_{3v}, C_{3v} \}$ & \qquad \qquad \quad (b) $\{C_{3}, C_{3} \}$
	\\
	 Space Group: $p3m1$ & \qquad \qquad  Space Group: $p3$
	
\end{tabular}

\begin{tabular}{ll}
		\includegraphics[width=3.25cm]{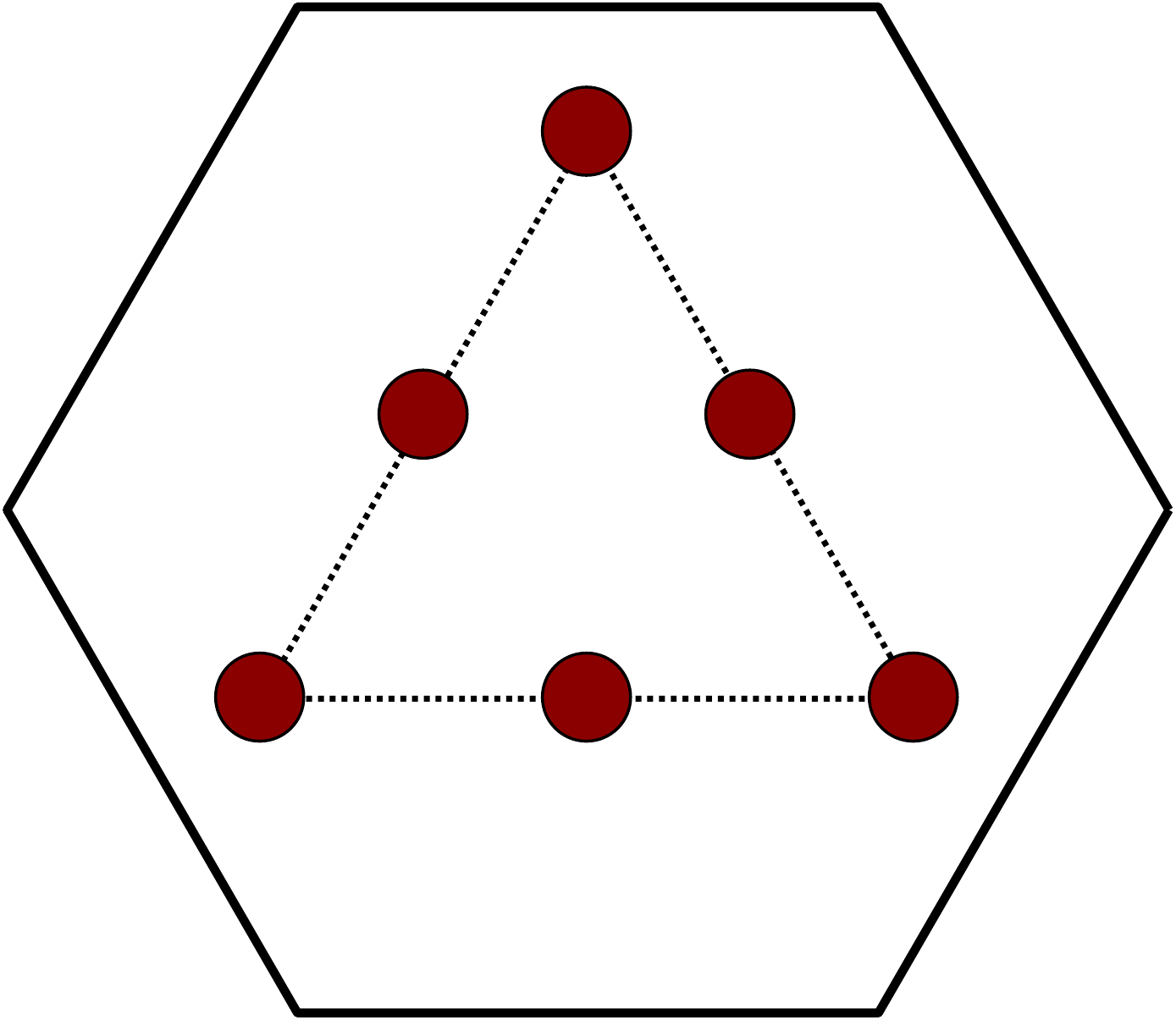} & 
	\hspace{2em} 
		\includegraphics[width=3.25cm]{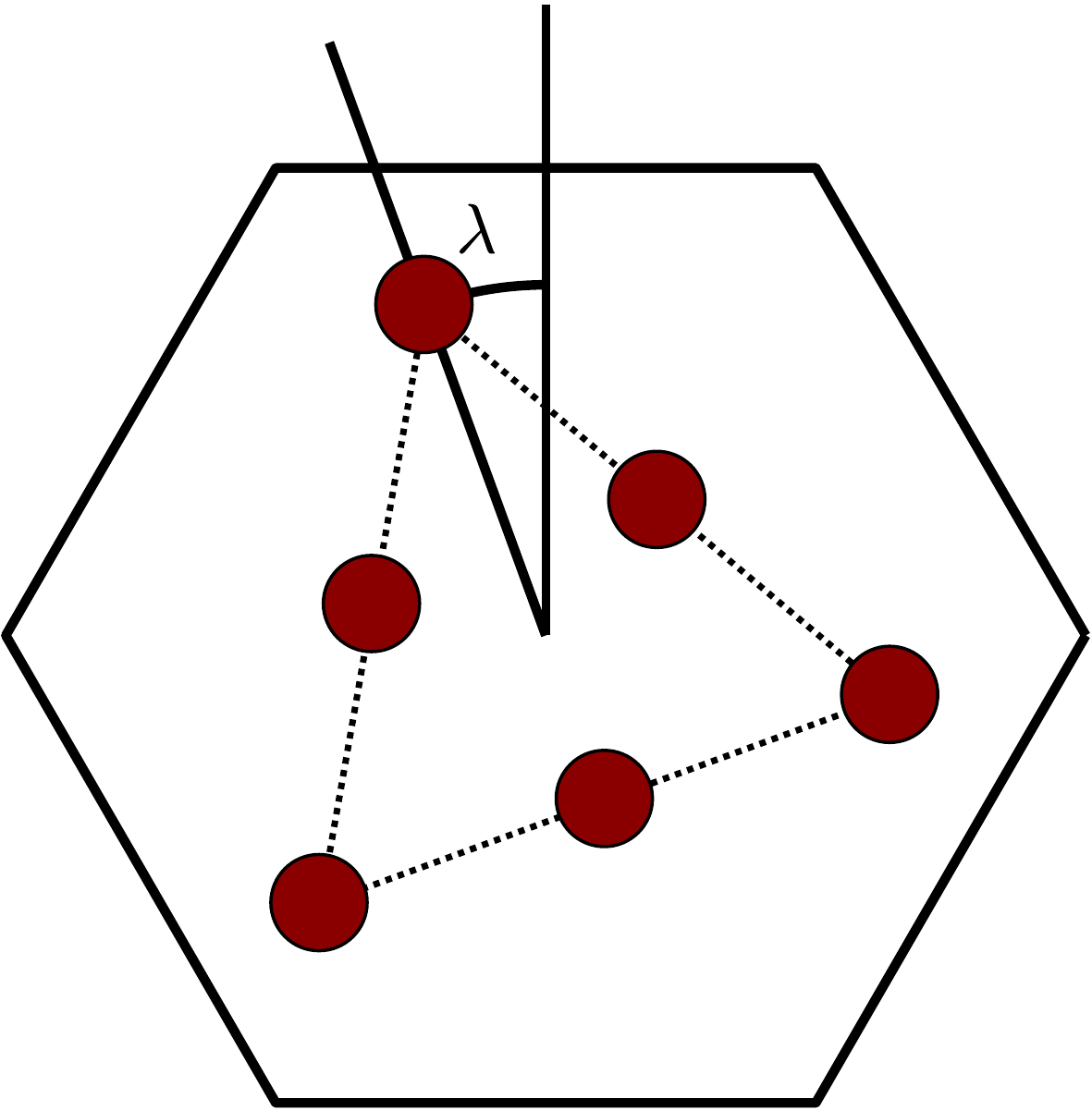} 
	\end{tabular}
		\includegraphics[width=8cm]{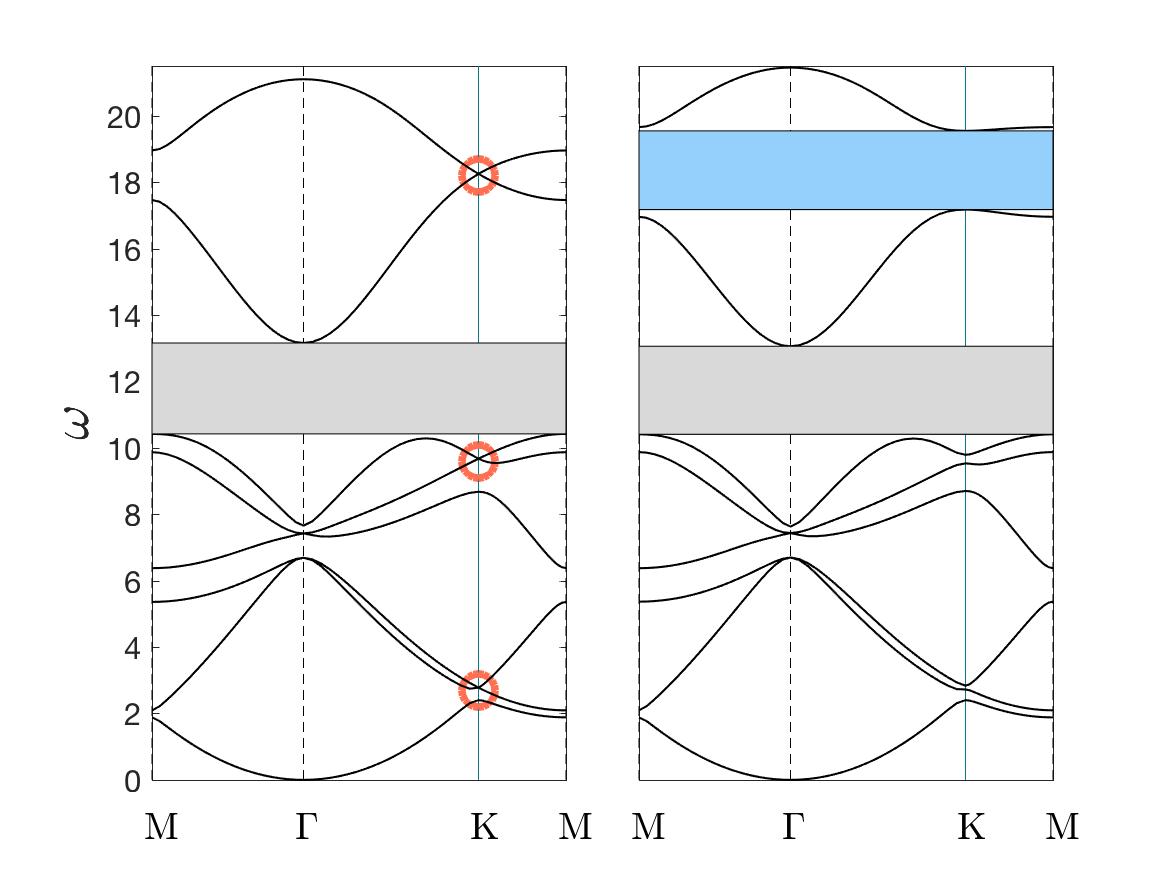}
	\caption{Case (ii) of table
          \ref{tab:hexagonal_DC}, generated by an angular perturbation of $\lambda=0.05$, showing
          the geometry, space groups and dispersion curves of (a)
          unperturbed and (b) perturbed cases: the norm of basis
          vectors is $2$ and distance between cell center and vertices of triangle is $0.8$. In (a) the
          Dirac points are circled and in (b) the new bandgap created
          by gapping the Dirac point is shaded blue.
}
\label{fig:C3v}
\end{figure}

For simplicity, we consider the honeycomb structure, comprised of the
AB sublattices, that, when unperturbed, has the system symmetries
$\{C_{6v}, C_{3v}\}$. The other two Dirac point gapping perturbations, used
in the subsequent section, arise from changing mass values and/or
positions; their perturbed Hamiltonian is obtained in a very similar
manner. 
Returning to the honeycomb structure, the effective Hamiltonian is given in \eqref{eq:Dirac_Hamiltonian} and the associated eigenvalues are shown in  \eqref{eq:linear_slope_K}. When perturbed, the mass term in \eqref{eq:ML_RHS} is expanded as, 
\beq
\begin{split}
M_{A, B} &= M_{av} +dM\left(x_{A, B}\right) = M_{av} \pm \alpha \Delta M, \\
M_{av} &= \frac{M_A + M_B}{2}, \quad \Delta M = \frac{|M_A - M_B|}{2},
\label{eq:mass_pert} \end{split} \eeq 
where, hereafter, we assume that $\Delta M$ is of the same order as $|\Delta \bkappa|$. The leading-order $M_{av}$ part is independent of the AB sublattices and retains the $C_{6v}$ 
symmetry of the unperturbed crystal. After substituting \eqref{eq:mass_pert} into \eqref{eq:expansion_gov} the bracketed mass term is altered to,
\beq
\sum_{{\bf N}, p} \delta\left({\bf x}-{\bf x}^{(p)}_{\bf N}\right) 
\left[ (\omega_{\bkappa_0}^2-\omega_{\bkappa}^2) M_{av} + \omega_{\bkappa_0}^2 dM\left(x_{A,B}\right) \right].
\eeq 
Consequently $M^{(p)}$ is replaced with, the $p$ independent term, $M_{av}$ up until \eqref{eq:Dirac_Hamiltonian}. The additional perturbative term, within the bracket, in \eqref{eq:reduced_linear_equation} is,
\beq
d\Lambda_{lj}=\sum_{p = A, B} \omega_{\bkappa_0}^2 dM\left(x_p\right) \psi^*_{l\bkappa_0}\left(\bx_p\right) \psi_{j\bkappa_0}\left(\bx_p\right).
\label{eq:pert_term}
\eeq 
If $M_A \neq M_B$ then the point group symmetry at the $KK^\prime$ valleys is reduced to $C_3$ hence the Dirac point is gapped. This is due to the breakdown in parity symmetry between the valleys in Fourier space and is reflected in the symmetry reduced counterpart to \eqref{eq:Lambda_matrix_elts_relationship},
\beq
d\Lambda_{lj}= -\sum_{m, n} D^*(\sigma_v)_{l, m} D(\sigma_v)_{n, j} d\Lambda_{mn}.
\label{eq:dLambda_matrix_elts_relationship}\eeq
Using this, and \eqref{eq:Dirac_Hamiltonian}, the perturbed effective Hamiltonian takes the form,
\beq
H^{(1)}_{\text{eff}} = H^{(1)}_{\text{unpert}} + \tau_z M_{K}\hat{\sigma}_y, \quad
M_{K} = \omega_{\bkappa_0}^2 \Delta M, 
\label{eq:pert_Ham_term}\eeq hence, due to the presence of $\tau_z$, $M_K =-M_{K^\prime}$. The corresponding eigenvalues for this effective Hamiltonian are,
\beq
(\omega_{\bkappa_0}^2-\omega_{\bkappa}^2) = \pm \sqrt{v^2_D |\Delta\bkappa|^2 + M^2_{K}}.
\label{eq:perturbed_eigenvalues}
\eeq 
This is precisely the form of the eigenvalues for the massive Dirac fermionic equation albeit, here, it is for a plate crystal. The, now, non-zero Dirac mass term in \eqref{eq:perturbed_eigenvalues} implies locally quadratic curvature. Following similar arguments to \cite{ochiai_photonic_2012} we   evaluate the Chern number as $C = C_K + C_{K^\prime}$ where $C_{K, K^\prime} = \text{sgn}\left(M_{K, K^\prime}\right)/2 = \pm \phi/2$;
the first relation arises because the greatest modal contribution emanates from the $KK^\prime$ valleys; the latter relation, with $\phi$, becomes evident after applying Stokes theorem to the Berry phase \eqref{eq:Berry_phase}. Due to the presence of the valley matrix index, $\tau_z$, in \eqref{eq:perturbed_eigenvalues} it follows that the Chern number vanishes; therefore a globally nontrivial topology is not permitted within the framework of our effective theory. However, the non-zero topological quantities at each of the valleys indicates that edge states, with limited backscattering, are possible as long as the hybridisation of the two valley modes is controlled.

\begin{figure} [h!]
\begin{tabular}{ll}
	\qquad (a) $\{C_{6}, C_{3} \}$ & \qquad \qquad \quad (b) $\{C_{3}, C_{3} \}$
	\\
	 Space Group: $p6$ & \qquad \qquad  Space Group: $p3$
	
\end{tabular}

\begin{tabular}{ll}
		\includegraphics[width=3.25cm]{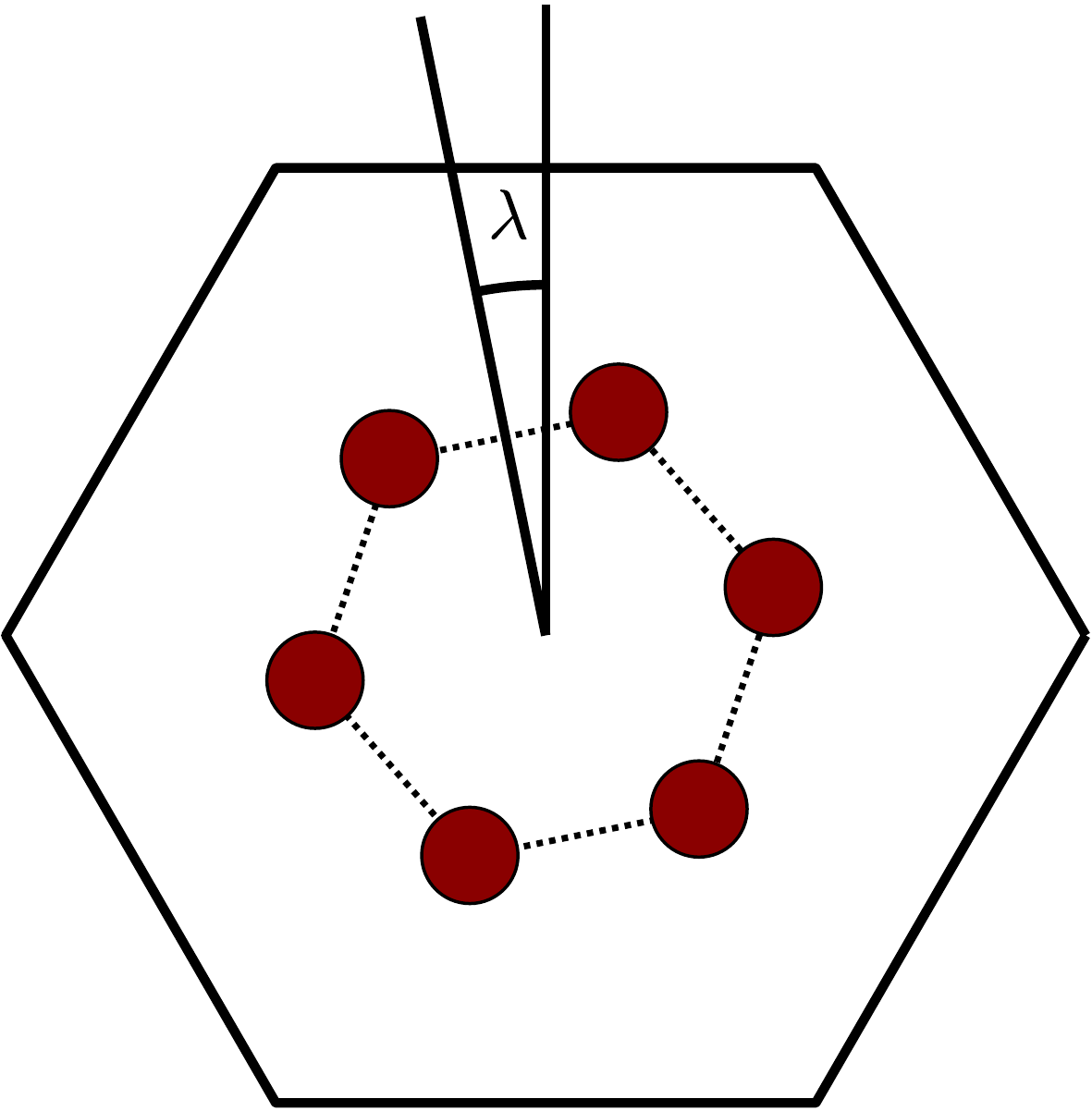} & 
	\hspace{2em} 
		\includegraphics[width=3.25cm]{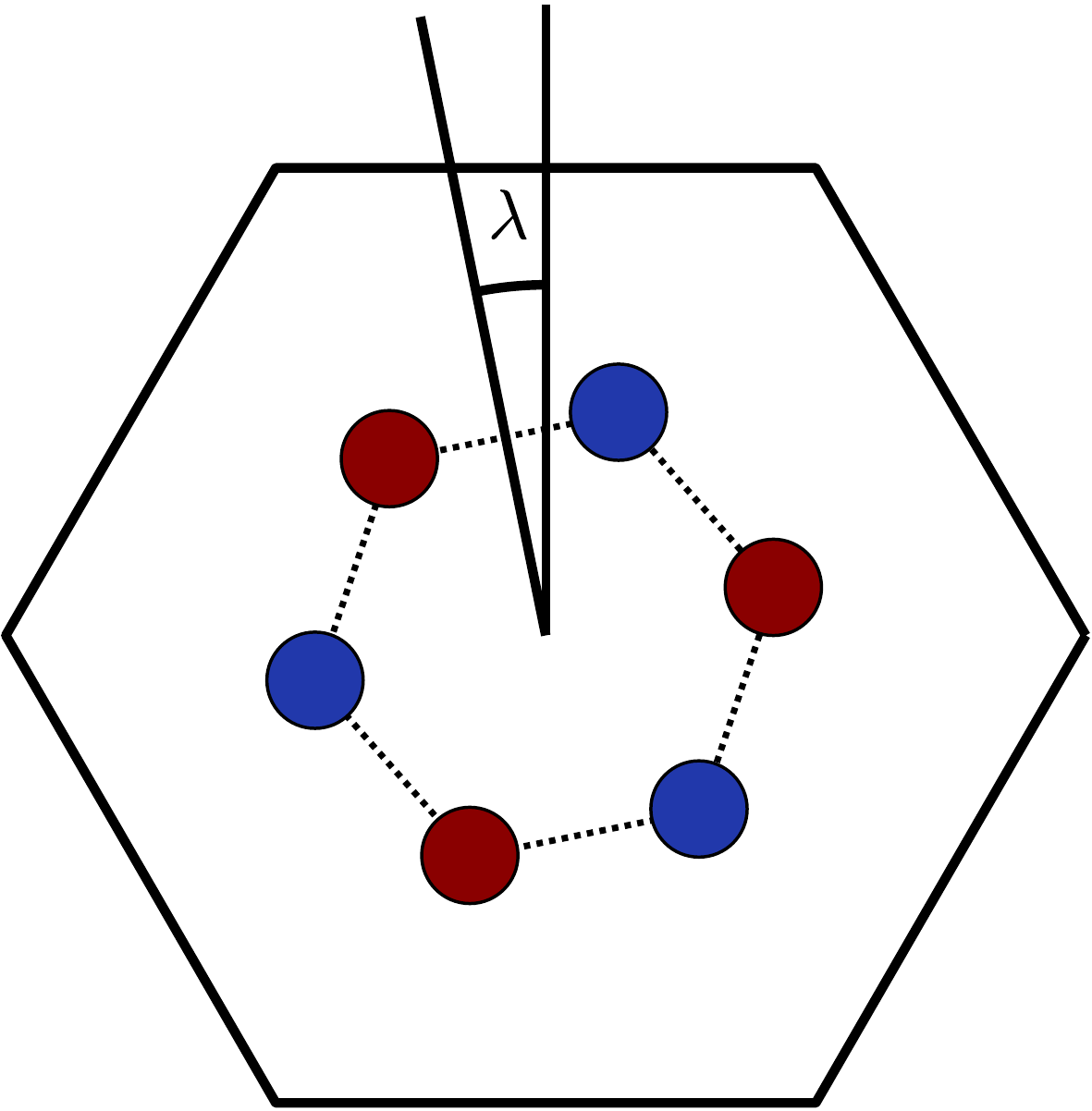} 
	\end{tabular}
		\includegraphics[width=8cm]{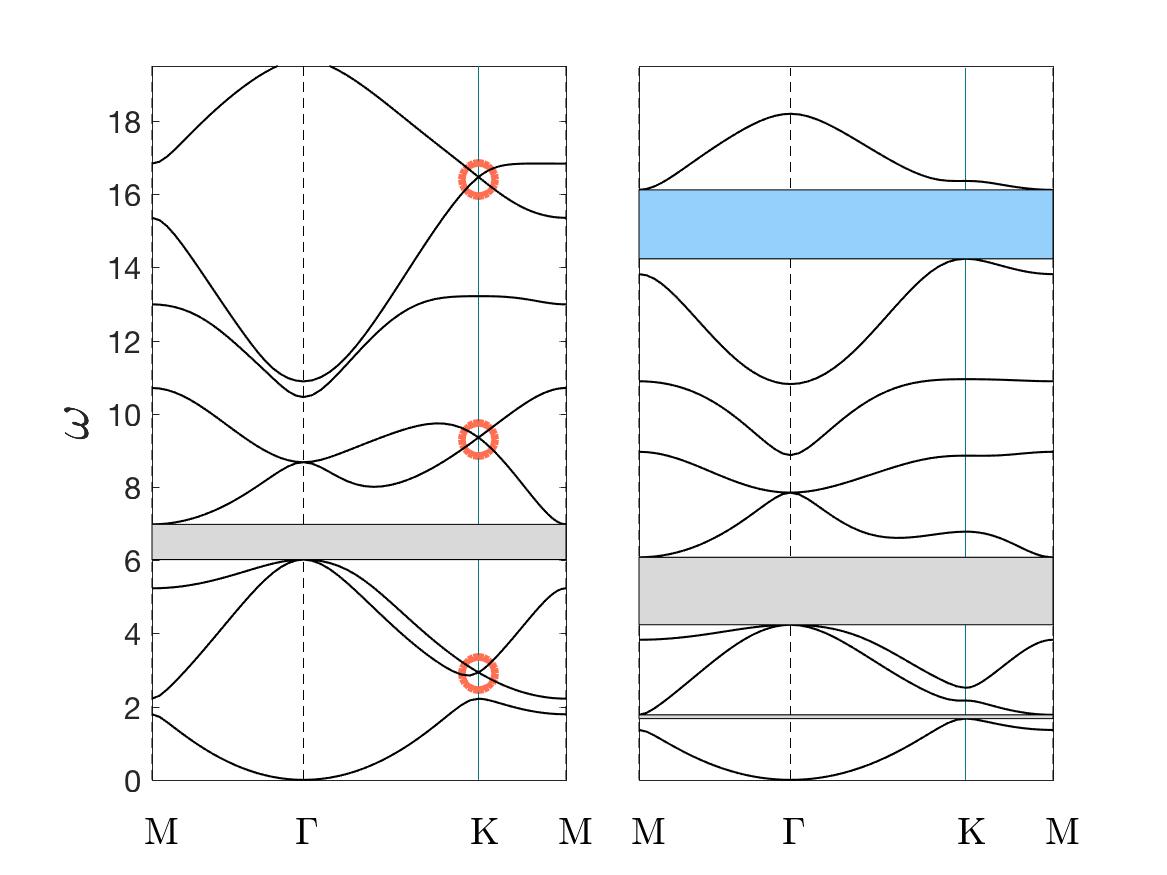}
	\caption{Case (iii) in table
          \ref{tab:hexagonal_DC}, generated by perturbing masses and geometry, showing
          the geometry, space groups and dispersion curves of (a)
          unperturbed and (b) perturbed cases: Angular deviation $\lambda$,
from cell symmetry axis, is is $0.15$; unperturbed mass value is $1$,
perturbed mass value is $2$. Norm of basis vectors is $2$ and distance
between cell center and masses is $0.5$. In (a) the
          Dirac points are circled and in (b) the new bandgap created
          by gapping the Dirac point is shaded blue. } 
\label{fig:C6}
\end{figure}


\section{Engineering topological valley states in plate crystals}
\label{sec:engineering}

\subsection{Symmetry reduction}

The three sets of symmetries,
table \ref{tab:hexagonal_DC}, give rise to deterministic
Dirac cones. The systematic reduction in symmetry of a plate
crystal will gap the Dirac points
thereby yielding a band gap. More specifically, for cases (i) and (ii), by reducing
$G_{KK^\prime}$ from $C_{3v}$ down to the symmetry set $C_3$ (or lower) we are assured
of a pair of non-degenerate quadratic curves at the $KK^\prime$
valleys. For case (i), if $G_{KK^\prime} = C_3$ then we would have to ensure 
that $G_{\Gamma}$ did not have six-fold symmetry. Case (iii) already has $G_{KK^\prime}=C_3$ and
the task is to lower the $G_{\Gamma}$ symmetry set by breaking inversion symmetry. 


Consider cases (i) and (ii)  of table \ref{tab:hexagonal_DC} which, for an unperturbed
system, have $\{C_{6v}, C_{3v} \}$ or $\{C_{3v}, C_{3v} \}$
symmetries; these two are reduced down to the symmetry set
$\{C_{3v},C_3\}$ or lower. The appropriate symmetry breaking, for
these cases, occurs when the mirror symmetry set $\{\sigma_v\}$ (and, for case (i), inversion symmetry as well) 
is removed in reciprocal space. 
Removing these mirror symmetries $\{\sigma_v\}$
c.f. Fig. \ref{fig:reflectional} is equivalent to the removal of $\{\sigma_d\}$ in physical space (see
Fig. \ref{fig:wavefield}). This is evident in the eigensolutions, shown in Fig.
\ref{fig:wavefield}, which demonstrate the symmetry breaking in physical 
space. The gapped Dirac points result in eigensolutions which have opposite chirality at the $KK^\prime$ valleys, 
Fig. \ref{fig:chirality}; this opposite pseudo-spin between the valleys is represented by $\tau_z$ in Eq. \eqref{eq:pert_Ham_term}.
A canonical example of this symmetry breaking occurs for the
honeycomb structure when the alternating sublattice mass values 
are made inequivalent; similar to 
Fig. \ref{fig:C6v}. Conversely, removing $\{\sigma_d\}$ symmetries in
Fourier space preserves the Dirac point, as $G_{KK^\prime}=C_{3v}$, and hence there
are no gapped states. For example, if 
the cellular structure of Fig. \ref{fig:C6v}(b) is rotated by $\pi/6$ the Dirac point will remain ungapped, despite inversion symmetry being broken, as the $\{\sigma_v\}$ symmetries remain. 

\begin{figure} 
\centering
\begin{tabular}{ll}
	(a) Case (ii) - $2\pi/3$ bend & (b) Case (ii) - $\pi/3$ bend
\end{tabular}	

\begin{tabular}{ll}	
	\includegraphics[scale=0.15]{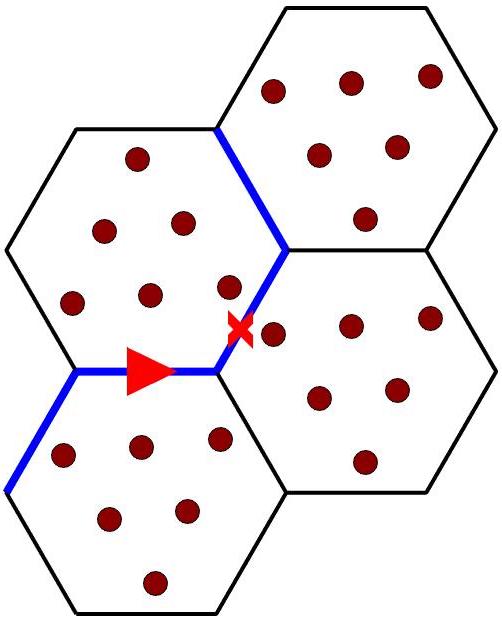} &
	\includegraphics[scale=0.15]{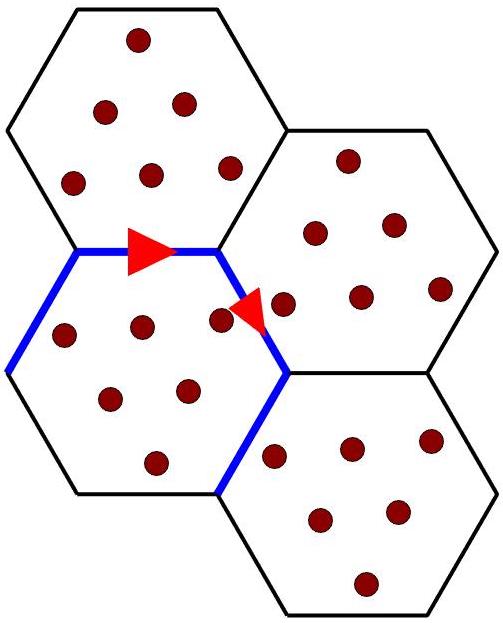}
\end{tabular}

\begin{tabular}{ll}
	(c) Case (i) - $2\pi/3$ bend & (d) Case (i) - $\pi/3$ bend
\end{tabular}	

\begin{tabular}{ll}	
	\includegraphics[scale=0.15]{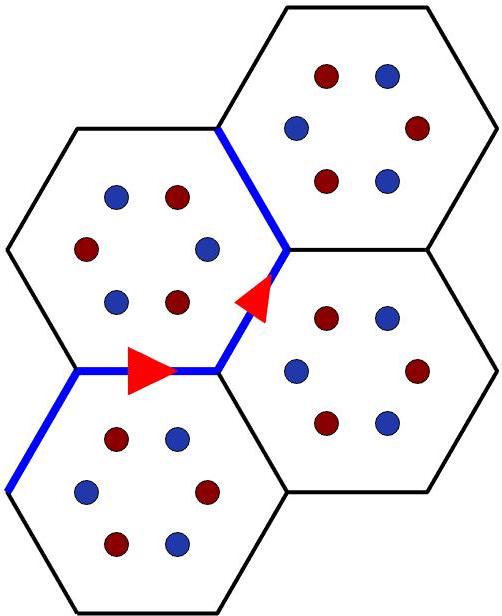} &
	\includegraphics[scale=0.15]{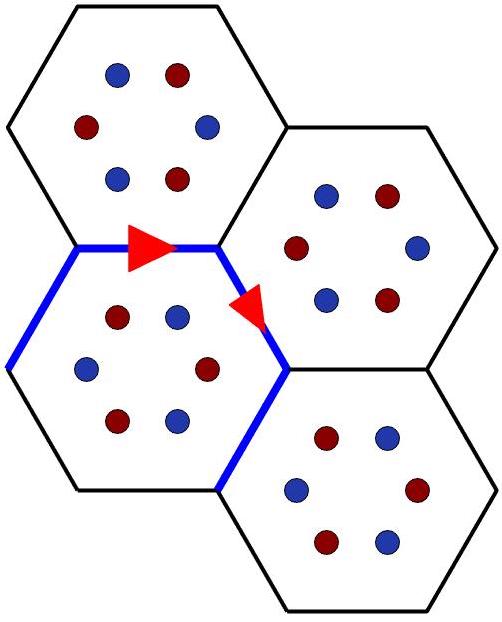}
\end{tabular}
\caption{The detail of the junction cells, showing their asymmetric edges, for the cases we consider. }
\label{fig:edges}
\end{figure}

Simple examples for the systematic reductions of all three symmetry
sets, and analysis of the
resulting topological states, are given in the subsequent sections. In
particular Fig. \ref{fig:C6v}(a) shows case (i), $\{C_{6v},C_{3v}\}$, 
reduced to $\{C_{3},C_3\}$ by the simple act of breaking reflectional
  symmetry through alternating the masses; case (ii)
  $\{C_{3v},C_{3v}\}$ has symmetry broken by rotating the inner
  triangular arrangement. Case (iii), with the unperturbed state as $\{C_6,C_3\}$, is slightly
different to the other two cases as $G_{KK^\prime}$ is already $C_3$  and
only supports a deterministic Dirac cone if 
$G_\Gamma = C_6$; hence the reduction of $G_\Gamma$ to $C_3$ is sufficient to gap the Dirac cone. Fig. \ref{fig:C6} 
shows the consequence of breaking
the $C_6$ symmetry at $\Gamma$, in this case alternating
the masses removes inversion symmetry.


\begin{figure} 
\begin{tabular}{ll}
\hspace{-2cm} (a) Orange medium & \qquad \qquad (b) Blue medium 
\end{tabular}

\begin{tabular}{ll}
		\includegraphics[scale=0.190]{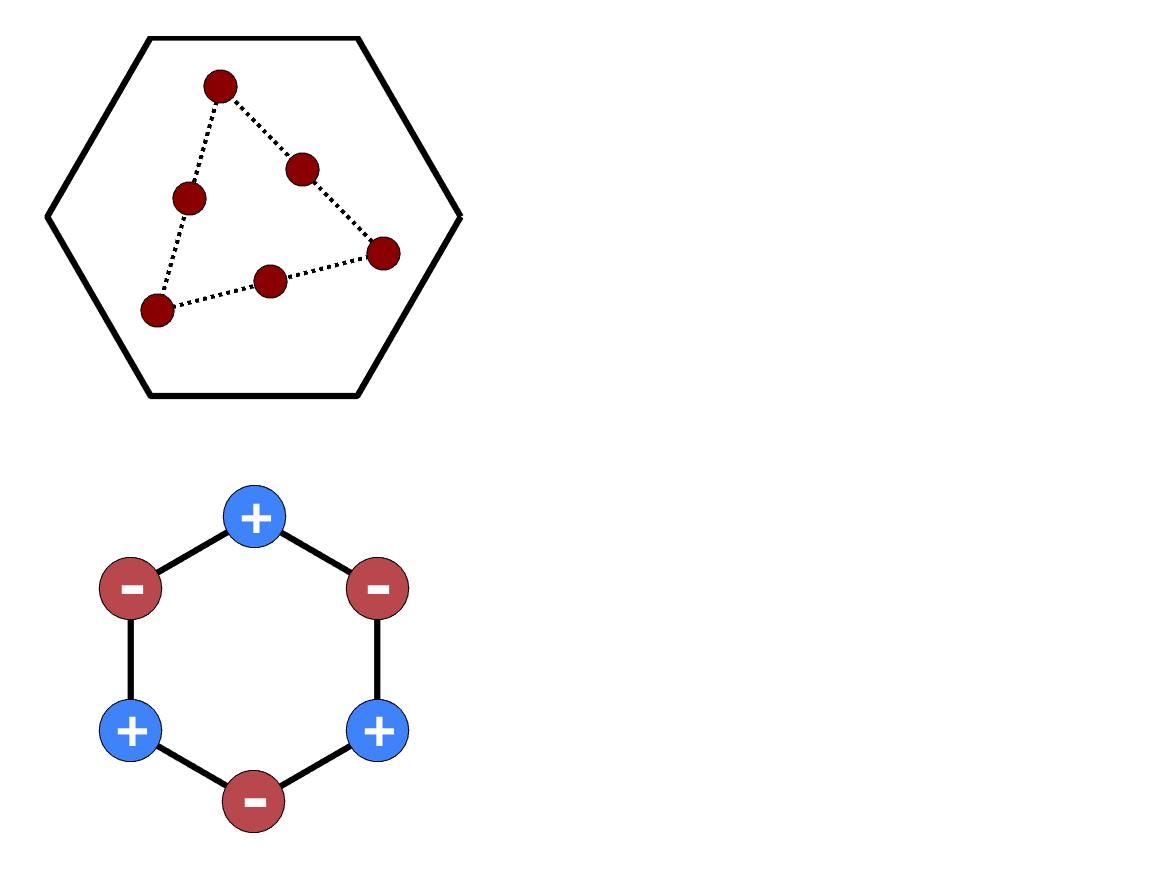} &
	\hspace{-4cm}	\includegraphics[scale=0.190]{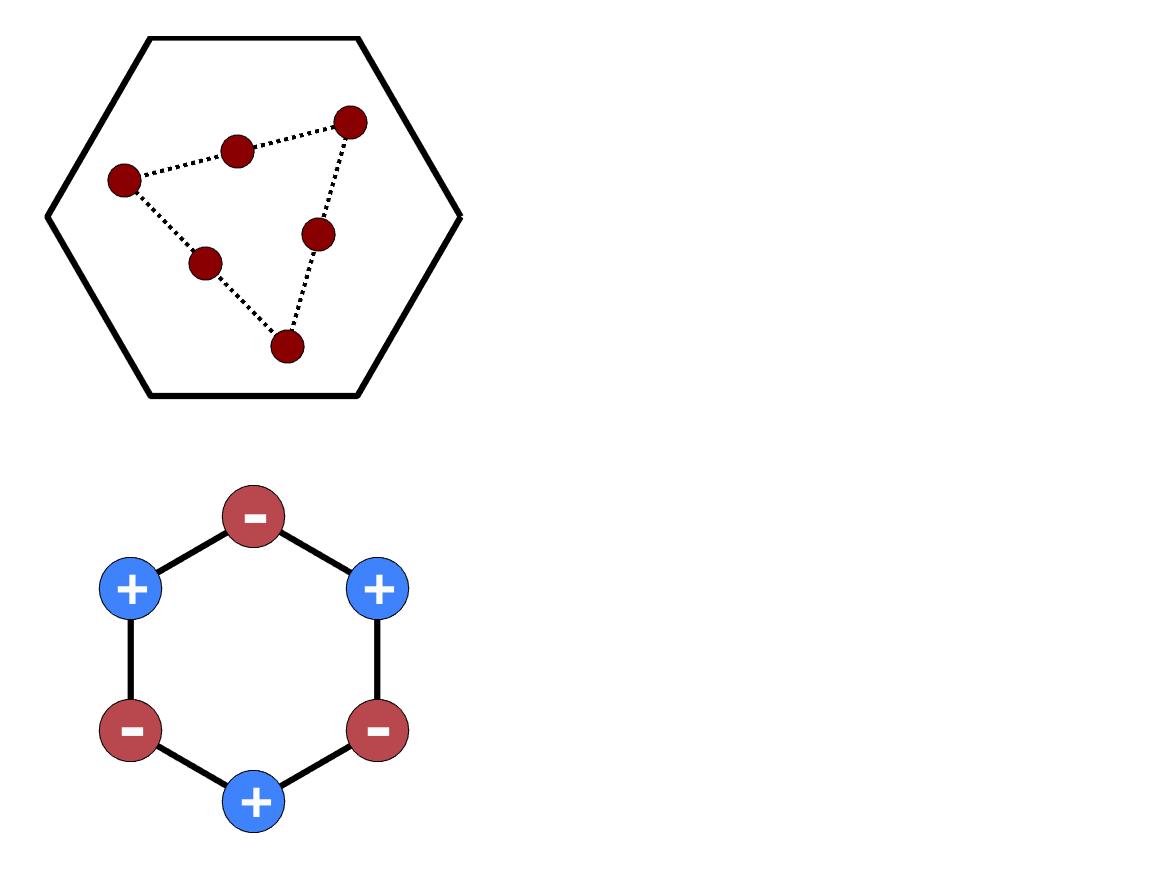}

	\end{tabular}
		(c) Ribbon dispersion curves and edge mode 
		\includegraphics[width=9cm]{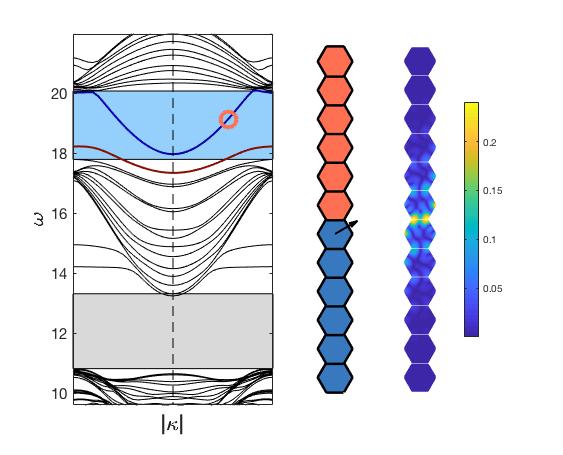}
	\caption{Plate zero-line mode. (a) and (b) showing the
          upper and lower media (case (ii) of Table
          \ref{tab:hexagonal_DC} as shown in Fig. \ref{fig:C3v}(b) for
        the upper medium and its rotation by $\pi/3$ for the lower
        medium), see Fig. \ref{fig:edges}(a), and the sign of the valley Chern numbers at the $KK^\prime$
        points. (c) shows the dispersion curves with the edge mode
        lying in the bandgap, $\omega = \{17.79, 20.07\}$, and at the highlighted frequency $\omega = 19.13$, we show the edge mode in physical
        space. 
}
\label{fig:C3v_ZLM}
\end{figure}

Figs \ref{fig:C6v}, \ref{fig:C3v}, \ref{fig:C6} show the bandstructures
for the cases of interest. The positions in reciprocal space
we choose deserve a note of detail regarding time-reversal invariant
systems. From the cell configurations shown in Figs
\ref{fig:C6v}, \ref{fig:C3v}, \ref{fig:C6} only the configuration in
Fig. \ref{fig:C6v}(a) reduces down to the IBZ, shown in Fig. \ref{fig:wigner}, with just the use of its spatial
symmetries. The presence of TRS permits a
further folding, on top of the spatial symmetries, thereby yielding an
IBZ having the $p6m$ space group IBZ.

\begin{figure} 
\begin{tabular}{ll}
\hspace{-1cm} (a) Orange medium & \qquad \qquad \qquad (b) Blue medium 
\end{tabular}

\begin{tabular}{ll}
		\includegraphics[scale=0.210]{Perturbed_C6v_WS_Rot.pdf} &
	\hspace{1cm}	\includegraphics[scale=0.210]{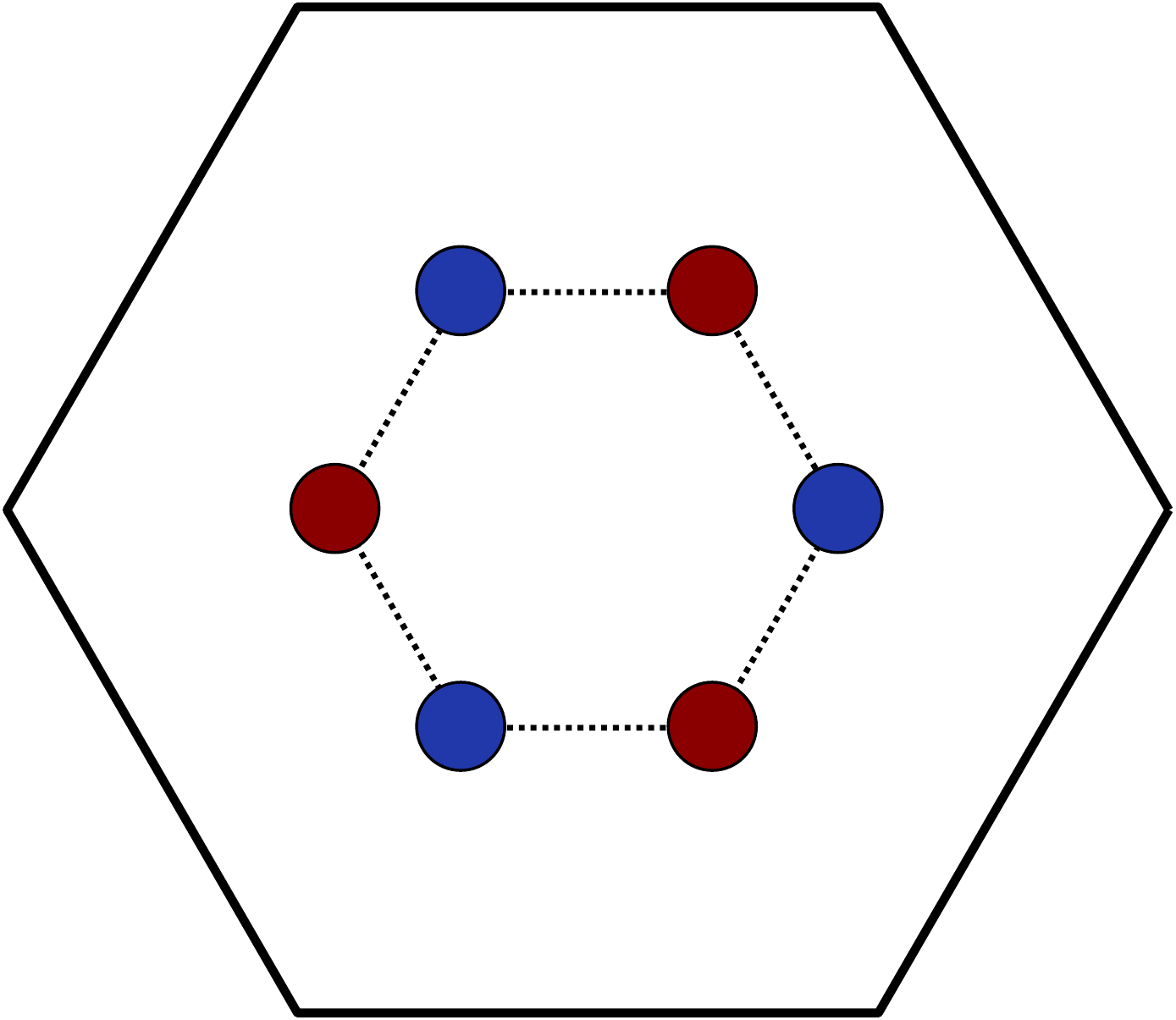}
	\end{tabular}
	
	(c) Ribbon dispersion curves and edge mode. 
		\includegraphics[width=8.5cm]{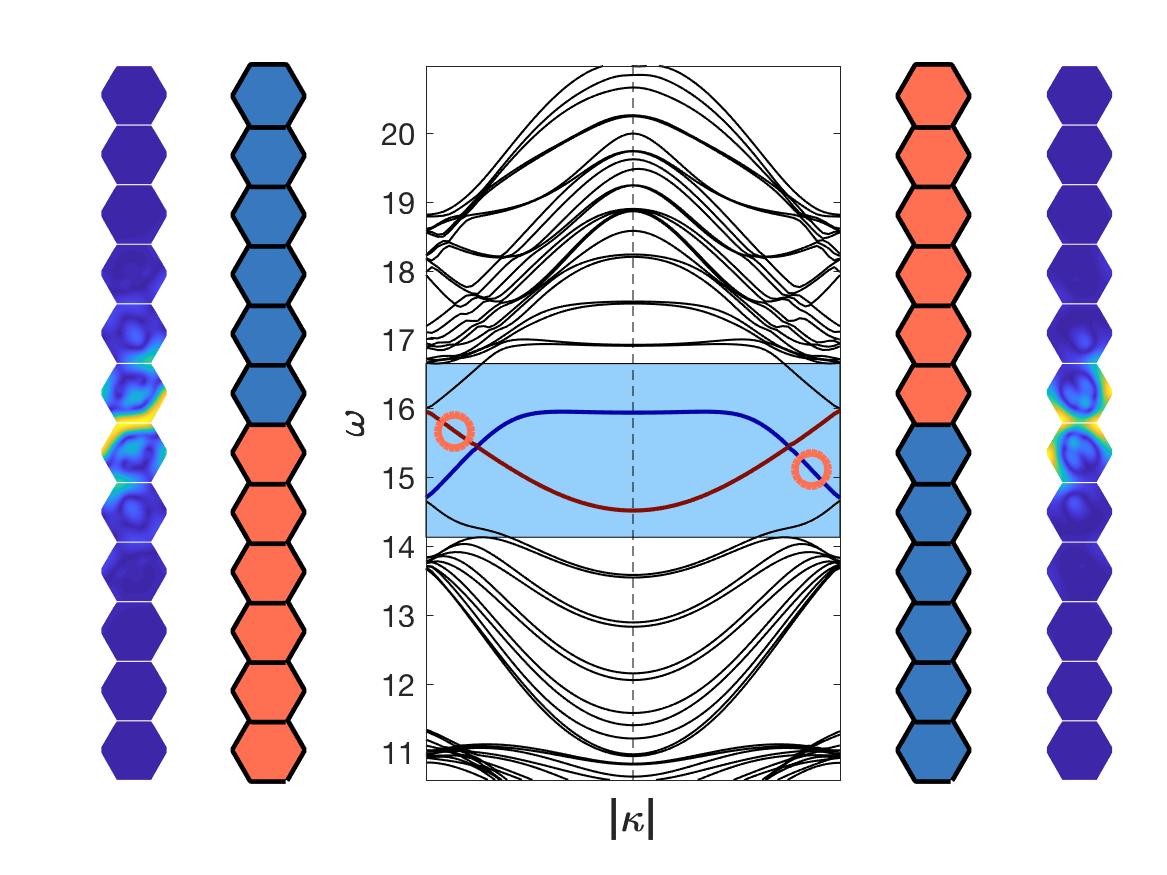}

	\caption{Plate zero-line mode. (a) and (b) showing the
          upper and lower media (case (i) of Table
          \ref{tab:hexagonal_DC} as shown in Fig. \ref{fig:C6v}(b) for
        the upper medium and its rotation by $\pi/3$ for the lower
        medium), see Fig. \ref{fig:edges}(b). (c) shows the dispersion curves with the edge modes
        lying in the bandgap, $\omega = \{14.13, 16.65\}$, and at the highlighted frequencies $\omega = 15.11, 15.67$, we show the edge modes in physical space.}
\label{fig:C6v_ZLM}
\end{figure}

Several papers have discussed how different
inclusion shapes affect the bandgap width \cite{min06a,wang07a,wang11a}. Although no
systematic rules can be drawn from these numerical studies several
factors appear to significantly impact the presence and width of bandgaps. \cite{wang01a} showed that the largest absolute photonic bandgap is
achieved by selecting an inclusion of the same symmetry as the cell.
 Other factors influencing the gap width are the orientation and size
 of the inclusions. 
 
\subsection{Plate zero-line modes}
The perturbation of the media has broken the six-fold symmetry whilst retaining three-fold symmetry. This results in 
asymmetric behaviour at the edges (and vertices) of the perturbed cellular structures. 
Due to the broken parity symmetry we now have inequivalent $KK^\prime$ valleys, and moreover the 
valley Chern numbers are opposite in sign ($\tau_z$ in Eq. \eqref{eq:pert_Ham_term}). 
Creating interfaces between two media that have opposite Chern numbers, at a particular valley,
will generate valley Hall edge states, \cite{xiao_valley-contrasting_2007}; these are aptly named zero-line modes (ZLMs) due to the opposite Chern numbers either side of the interface. 
A convenient approach for creating such media is to take advantage of the lack of six-fold symmetry and join 
one perturbed medium to its $\pi/3$ rotated twin. A benefit of these valley Hall modes is that we have {\it a priori} knowledge of how to construct the two adjoining media (sharing a bandgap) 
such that we are guaranteed broadband edge modes.

To compute the ZLMs we adapt the plane wave expansion method,
\eqref{eq:transformed}, by extending it to a finite, long, ribbon of hexagons (as
shown in the inset to Fig. \ref{fig:C3v_ZLM}). We need only consider Bloch conditions
in the direction indicated by the arrow shown in the hexagonal ribbon
of Fig. \ref{fig:C3v_ZLM}. Numerically, we consider a long ribbon and extract
solutions that decay exponentially; we apply periodic conditions at
the top/bottom of the ribbon and convergence is checked by mode doubling
and extending the ribbon.  

The three-fold symmetry of the perturbed structures, resulting in asymmetric edges, yields two distinct interfaces between two topologically distinct media; these distinct edges are shown in Fig. \ref{fig:edges}. ZLMs are shown for two different geometrical cases; namely, cases (i) and (ii). 
For the latter, we obtain the edge modes shown in Fig. \ref{fig:C3v_ZLM}; the broadband mode pertains 
to the orange cell over the blue cell, whilst the narrow band mode, below it, comes from flipping the cells i.e. the blue over orange; the different interfaces between the media coincide with the two distinct edges. The limited 
overlap between the two modes implies that for a large frequency range these modes do not couple into each other. This property will be used for filtering waves in section \ref{sec:bends}.

In contrast, Fig. \ref{fig:C6v_ZLM} for case (i) has two overlapping broadband modes. Unlike case (ii) both the 
edge modes for orange over blue (and vice versa) exist over a simultaneous frequency range. The distinction
in the edges is reflected in the different modal patterns shown in Fig. \ref{fig:C6v_ZLM}. The 
differences, between cases (i) and (ii), emerge from the degree of asymmetry of the cell's edges.
The large distance between the centroid and the vertices of the triangular inclusion for case (ii) results in
an almost effective barrier to wave transport (see Fig. \ref{fig:edges} (a) and (b)); this is unlike 
case (i) (Fig. \ref{fig:edges} (c) and (d)).


 \subsection{Transport around sharp edges}
\label{sec:bends}
To illustrate the consequences of particular cellular structures, we now examine 
the propagative behaviour around gentle and sharp bends of angles $2\pi/3, \pi/3$ respectively. Our simulations use the explicit Green's function to create a linear system 
 that is rapidly evaluated as in \cite{evans07a,torrent13a}. 

\begin{figure} [htb!]
\centering
\begin{tabular}{lll}
\hspace{1.5cm} (a) & \hspace{1.5cm}(b) & \hspace{1.5cm} (c) \\
 	\includegraphics[scale=0.400]{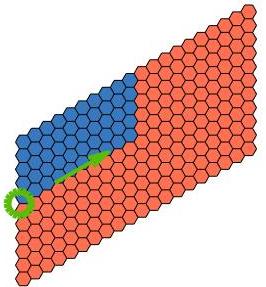} &
 		\includegraphics[scale=0.370]{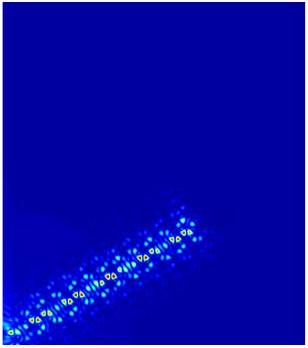} &
 		\includegraphics[scale=0.250]{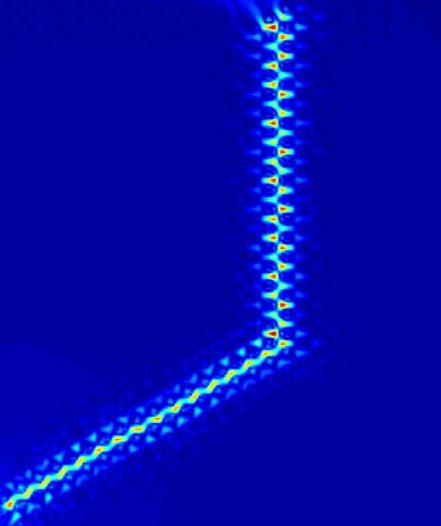}	
\end{tabular} 

\caption{ZLM transport around a gentle bend ($2\pi/3$ angle), schematic in (a). Panels (b) and (c) show the absolute value of displacements for case (ii) $C_{3v}$, and case (i) $C_{6v}$ respectively. The detail of the ZLM pre- and post-bend in (c) is of note.
$1890$ cells were used, excitation frequency for $C_{3v}$ case was $\omega = 19.53$, and $C_{6v}$ case $\omega = 15.19$.}
\label{fig:2piover3}
\end{figure}

Initially, we  analyse the gentle bend, Fig. \ref{fig:2piover3}(a), and we launch a 
ZLM at the leftmost interface towards the bend. Notably the leftmost interface
differs from the post-bend interface; Fig. \ref{fig:edges}(a) and (c) show a close-up
of the four cells at the bend. The edge pre- and post-bend are clearly different, 
the pre-bend interface is identical to the leftmost interface. This property 
explains the different results obtained for cases (i) and (ii), shown in Fig. \ref{fig:2piover3}(c) and (b),
respectively. For case (ii) the single broadband ZLM cannot mode convert around the bend as there is no
overlapping ZLM to couple into, Fig. \ref{fig:edges}(a). Conversely, for case (i) the broadband
ZLMs, that lie in a simultaneous range, allow for perfect coupling from one ZLM into the other, Fig.
\ref{fig:edges}(c); this is demonstrated by the distinct modal patterns pre- and post-bend, Fig. \ref{fig:2piover3}(c). The exactitude of our numerical solutions allows for 
visual clarity of the edge modes and hence easier interpretation of the underlying physical mechanisms.

\begin{figure} [htb!]
\centering
\begin{tabular}{lll}
\hspace{1.5cm} (a) & \hspace{1.5cm}(b) & \hspace{1.5cm} (c) \\
 	\includegraphics[scale=0.400]{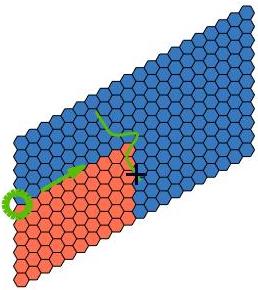} &
 		\includegraphics[scale=0.340]{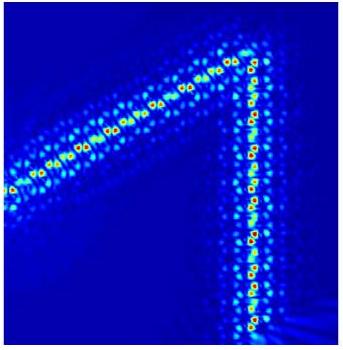} &
 		\includegraphics[scale=0.340]{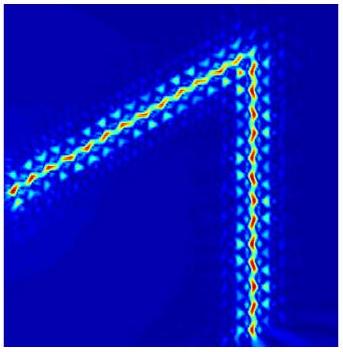}	
\end{tabular} 
\caption{ZLM transport around a sharp bend ($\pi/3$ angle), schematic in (a). Panels (b) and (c) show the absolute value of displacements for case (ii) $C_{3v}$, and case (i) $C_{6v}$ respectively. The details of the ZLM pre- and post-bends are of note. 
$1890$ cells were used, excitation frequency for $C_{3v}$ case was $\omega = 19.18$, and $C_{6v}$ case $\omega = 15.09$.}
\label{fig:piover3}
\end{figure}

Turning our attention to the sharp bend, Fig. \ref{fig:piover3}; from Fig. \ref{fig:edges}(b) and (d), along with the discussion in the previous paragraph, one 
would naively expect similar behaviour to that of the gentle bend. However this is not the case as 
the zigzag edges pre- and post-bend are identical; this is more easily seen from the scattering Figs. \ref{fig:piover3} rather than the cellular arrangement, Fig. \ref{fig:edges} (b) and (d).
Additionally a new phenomena, tunnelling, plays a role. These properties are numerically indicated in Fig. \ref{fig:piover3}, in particular note 
how the clarity of the modes allows us to see  with ease that the modal patterns pre- and post-bend, are identical, for both geometrical cases. 

\onecolumngrid

\twocolumngrid

\onecolumngrid

\twocolumngrid

\section{Concluding remarks}
\label{sec:conclusion}

 We have shown, using group and {\bf k}$\cdot${\bf p} theory, a first-principles approach to 
 creating plate topological valley modes for all two-dimensional hexagonal lattices.
 For the elastic plate crystal, this 
 allows for there to be a direct bridge between the quantum and continuum
 mechanical worlds. Group theory allowed us to identify three distinct geometrical cases having symmetry 
 induced Dirac cones; the correct breaking of parity symmetry revealed 
nontrivial  band-gaps in which broadband edge modes reside. Given such edge modes 
it is natural to consider the transport of energy around bends in partitioned media. 
It becomes clear when doing so that the details of the junction cells, in the vicinity of the bend, are 
significant. One important issue that we have not considered here is
that of disorder, \cite{jin18a} consider this and it would be
interesting to explore this carefully for the ZLMs presented here. 
The asymmetry of the edges at cells, between the topologically distinct media, 
indicates two different mechanisms of energy transport around bends; the success of the strategy we employ is seen through the clarity of states demonstrated numerically.

Armed with this knowledge of how to design really clean and sharp edge states, and the underlying principles at the junction cells, this will motivate the design of more efficient interfacial waveguides, topological networks and energy filters, using the different cases we have identified: Experimental verification of these results in encouraged. 

\section*{Acknowledgements}

The authors thank the EPSRC for their support through grant
{EP/L024926/1} and R.V.C acknowledges the support of a Leverhulme Trust
Research Fellowship. 

\bibliographystyle{apsrev4-1}

\begin{thebibliography}{54}%
\makeatletter
\providecommand \@ifxundefined [1]{%
 \@ifx{#1\undefined}
}%
\providecommand \@ifnum [1]{%
 \ifnum #1\expandafter \@firstoftwo
 \else \expandafter \@secondoftwo
 \fi
}%
\providecommand \@ifx [1]{%
 \ifx #1\expandafter \@firstoftwo
 \else \expandafter \@secondoftwo
 \fi
}%
\providecommand \natexlab [1]{#1}%
\providecommand \enquote  [1]{``#1''}%
\providecommand \bibnamefont  [1]{#1}%
\providecommand \bibfnamefont [1]{#1}%
\providecommand \citenamefont [1]{#1}%
\providecommand \href@noop [0]{\@secondoftwo}%
\providecommand \href [0]{\begingroup \@sanitize@url \@href}%
\providecommand \@href[1]{\@@startlink{#1}\@@href}%
\providecommand \@@href[1]{\endgroup#1\@@endlink}%
\providecommand \@sanitize@url [0]{\catcode `\\12\catcode `\$12\catcode
  `\&12\catcode `\#12\catcode `\^12\catcode `\_12\catcode `\%12\relax}%
\providecommand \@@startlink[1]{}%
\providecommand \@@endlink[0]{}%
\providecommand \url  [0]{\begingroup\@sanitize@url \@url }%
\providecommand \@url [1]{\endgroup\@href {#1}{\urlprefix }}%
\providecommand \urlprefix  [0]{URL }%
\providecommand \Eprint [0]{\href }%
\providecommand \doibase [0]{http://dx.doi.org/}%
\providecommand \selectlanguage [0]{\@gobble}%
\providecommand \bibinfo  [0]{\@secondoftwo}%
\providecommand \bibfield  [0]{\@secondoftwo}%
\providecommand \translation [1]{[#1]}%
\providecommand \BibitemOpen [0]{}%
\providecommand \bibitemStop [0]{}%
\providecommand \bibitemNoStop [0]{.\EOS\space}%
\providecommand \EOS [0]{\spacefactor3000\relax}%
\providecommand \BibitemShut  [1]{\csname bibitem#1\endcsname}%
\let\auto@bib@innerbib\@empty
\bibitem [{\citenamefont {Kane}\ and\ \citenamefont
  {Mele}(2005)}]{Kane_Mele_2005}%
  \BibitemOpen
  \bibfield  {author} {\bibinfo {author} {\bibfnamefont {C.~L.}\ \bibnamefont
  {Kane}}\ and\ \bibinfo {author} {\bibfnamefont {E.~J.}\ \bibnamefont
  {Mele}},\ }\href@noop {} {\bibfield  {journal} {\bibinfo  {journal} {Phys.
  Rev. Lett.}\ }\textbf {\bibinfo {volume} {95}},\ \bibinfo {pages} {146802}
  (\bibinfo {year} {2005})}\BibitemShut {NoStop}%
\bibitem [{\citenamefont {Hasan}\ and\ \citenamefont {Kane}(2010)}]{hasan10a}%
  \BibitemOpen
  \bibfield  {author} {\bibinfo {author} {\bibfnamefont {M.~Z.}\ \bibnamefont
  {Hasan}}\ and\ \bibinfo {author} {\bibfnamefont {C.~L.}\ \bibnamefont
  {Kane}},\ }\href@noop {} {\bibfield  {journal} {\bibinfo  {journal} {Rev.
  Mod. Phys.}\ }\textbf {\bibinfo {volume} {82}},\ \bibinfo {pages} {3045}
  (\bibinfo {year} {2010})}\BibitemShut {NoStop}%
\bibitem [{\citenamefont {Ren}\ \emph {et~al.}(2016)\citenamefont {Ren},
  \citenamefont {Qiao},\ and\ \citenamefont {Niu}}]{Ren_2016}%
  \BibitemOpen
  \bibfield  {author} {\bibinfo {author} {\bibfnamefont {Y.}~\bibnamefont
  {Ren}}, \bibinfo {author} {\bibfnamefont {Z.}~\bibnamefont {Qiao}}, \ and\
  \bibinfo {author} {\bibfnamefont {Q.}~\bibnamefont {Niu}},\ }\href@noop {}
  {\bibfield  {journal} {\bibinfo  {journal} {Rep. Prog. Phys.}\ }\textbf
  {\bibinfo {volume} {79}},\ \bibinfo {pages} {066501} (\bibinfo {year}
  {2016})}\BibitemShut {NoStop}%
\bibitem [{\citenamefont {Lu}\ \emph {et~al.}(2014{\natexlab{a}})\citenamefont
  {Lu}, \citenamefont {Joannopoulos},\ and\ \citenamefont {Soljacic}}]{lu14a}%
  \BibitemOpen
  \bibfield  {author} {\bibinfo {author} {\bibfnamefont {L.}~\bibnamefont
  {Lu}}, \bibinfo {author} {\bibfnamefont {J.~D.}\ \bibnamefont
  {Joannopoulos}}, \ and\ \bibinfo {author} {\bibfnamefont {M.}~\bibnamefont
  {Soljacic}},\ }\href@noop {} {\bibfield  {journal} {\bibinfo  {journal}
  {Nature Photonics}\ }\textbf {\bibinfo {volume} {8}},\ \bibinfo {pages} {821}
  (\bibinfo {year} {2014}{\natexlab{a}})}\BibitemShut {NoStop}%
\bibitem [{\citenamefont {Khanikaev}\ and\ \citenamefont
  {Shvets}(2017)}]{khanikaev_two-dimensional_2017}%
  \BibitemOpen
  \bibfield  {author} {\bibinfo {author} {\bibfnamefont {A.~B.}\ \bibnamefont
  {Khanikaev}}\ and\ \bibinfo {author} {\bibfnamefont {G.}~\bibnamefont
  {Shvets}},\ }\href {\doibase 10.1038/s41566-017-0048-5} {\bibfield  {journal}
  {\bibinfo  {journal} {Nat. Photonics}\ }\textbf {\bibinfo {volume} {11}},\
  \bibinfo {pages} {763} (\bibinfo {year} {2017})}\BibitemShut {NoStop}%
\bibitem [{\citenamefont {Dresselhaus}\ \emph {et~al.}(2008)\citenamefont
  {Dresselhaus}, \citenamefont {Dresselhaus},\ and\ \citenamefont
  {Jorio}}]{dresselhaus08a}%
  \BibitemOpen
  \bibfield  {author} {\bibinfo {author} {\bibfnamefont {M.~S.}\ \bibnamefont
  {Dresselhaus}}, \bibinfo {author} {\bibfnamefont {G.}~\bibnamefont
  {Dresselhaus}}, \ and\ \bibinfo {author} {\bibfnamefont {A.}~\bibnamefont
  {Jorio}},\ }\href@noop {} {\emph {\bibinfo {title} {Group theory: application
  to the physics of condensed matter}}}\ (\bibinfo  {publisher}
  {Springer-Verlag},\ \bibinfo {year} {2008})\BibitemShut {NoStop}%
\bibitem [{\citenamefont {Inui}\ \emph {et~al.}(1990)\citenamefont {Inui},
  \citenamefont {Tanabe},\ and\ \citenamefont {Onodera}}]{inui90a}%
  \BibitemOpen
  \bibfield  {author} {\bibinfo {author} {\bibfnamefont {T.}~\bibnamefont
  {Inui}}, \bibinfo {author} {\bibfnamefont {Y.}~\bibnamefont {Tanabe}}, \ and\
  \bibinfo {author} {\bibfnamefont {Y.}~\bibnamefont {Onodera}},\ }\href@noop
  {} {\emph {\bibinfo {title} {Group Theory and Its Applications in Physics}}}\
  (\bibinfo  {publisher} {Springer-Verlag},\ \bibinfo {year}
  {1990})\BibitemShut {NoStop}%
\bibitem [{\citenamefont {Dai}\ \emph {et~al.}(2017)\citenamefont {Dai},
  \citenamefont {Liu}, \citenamefont {Jiao}, \citenamefont {Xia},\ and\
  \citenamefont {Yu}}]{Dai_2017}%
  \BibitemOpen
  \bibfield  {author} {\bibinfo {author} {\bibfnamefont {H.}~\bibnamefont
  {Dai}}, \bibinfo {author} {\bibfnamefont {T.}~\bibnamefont {Liu}}, \bibinfo
  {author} {\bibfnamefont {J.}~\bibnamefont {Jiao}}, \bibinfo {author}
  {\bibfnamefont {B.}~\bibnamefont {Xia}}, \ and\ \bibinfo {author}
  {\bibfnamefont {D.}~\bibnamefont {Yu}},\ }\href@noop {} {\bibfield  {journal}
  {\bibinfo  {journal} {Journal of Applied Physics}\ } (\bibinfo {year}
  {2017})}\BibitemShut {NoStop}%
\bibitem [{\citenamefont {Lu}\ \emph {et~al.}(2014{\natexlab{b}})\citenamefont
  {Lu}, \citenamefont {Chunyin~Qiu}, \citenamefont {Ye}, \citenamefont {Ke},\
  and\ \citenamefont {Liu}}]{Lu_2014}%
  \BibitemOpen
  \bibfield  {author} {\bibinfo {author} {\bibfnamefont {J.}~\bibnamefont
  {Lu}}, \bibinfo {author} {\bibfnamefont {S.~X.}\ \bibnamefont {Chunyin~Qiu}},
  \bibinfo {author} {\bibfnamefont {Y.}~\bibnamefont {Ye}}, \bibinfo {author}
  {\bibfnamefont {M.}~\bibnamefont {Ke}}, \ and\ \bibinfo {author}
  {\bibfnamefont {Z.}~\bibnamefont {Liu}},\ }\href@noop {} {\bibfield
  {journal} {\bibinfo  {journal} {Phys. Rev. B}\ } (\bibinfo {year}
  {2014}{\natexlab{b}})}\BibitemShut {NoStop}%
\bibitem [{\citenamefont {Mc{P}hedran}\ \emph {et~al.}(2009)\citenamefont
  {Mc{P}hedran}, \citenamefont {Movchan},\ and\ \citenamefont
  {Movchan}}]{mcphedran09a}%
  \BibitemOpen
  \bibfield  {author} {\bibinfo {author} {\bibfnamefont {R.~C.}\ \bibnamefont
  {Mc{P}hedran}}, \bibinfo {author} {\bibfnamefont {A.~B.}\ \bibnamefont
  {Movchan}}, \ and\ \bibinfo {author} {\bibfnamefont {N.~V.}\ \bibnamefont
  {Movchan}},\ }\href@noop {} {\bibfield  {journal} {\bibinfo  {journal} {Mech.
  Mat.}\ }\textbf {\bibinfo {volume} {41}},\ \bibinfo {pages} {356} (\bibinfo
  {year} {2009})}\BibitemShut {NoStop}%
\bibitem [{\citenamefont {Landau}\ and\ \citenamefont
  {Lifshitz}(1970)}]{landau70a}%
  \BibitemOpen
  \bibfield  {author} {\bibinfo {author} {\bibfnamefont {L.~D.}\ \bibnamefont
  {Landau}}\ and\ \bibinfo {author} {\bibfnamefont {E.~M.}\ \bibnamefont
  {Lifshitz}},\ }\href@noop {} {\emph {\bibinfo {title} {Theory of
  elasticity}}},\ \bibinfo {edition} {2nd}\ ed.\ (\bibinfo  {publisher}
  {Pergamon Press},\ \bibinfo {year} {1970})\BibitemShut {NoStop}%
\bibitem [{\citenamefont {Graff}(1975)}]{graff75a}%
  \BibitemOpen
  \bibfield  {author} {\bibinfo {author} {\bibfnamefont {K.~F.}\ \bibnamefont
  {Graff}},\ }\href@noop {} {\emph {\bibinfo {title} {Wave motion in elastic
  solids}}}\ (\bibinfo  {publisher} {Oxford University Press},\ \bibinfo {year}
  {1975})\BibitemShut {NoStop}%
\bibitem [{\citenamefont {Lefebvre}\ \emph {et~al.}(2017)\citenamefont
  {Lefebvre}, \citenamefont {Antonakakis}, \citenamefont {Achaoui},
  \citenamefont {Craster}, \citenamefont {Guenneau},\ and\ \citenamefont
  {Sebbah}}]{lefebvre17a}%
  \BibitemOpen
  \bibfield  {author} {\bibinfo {author} {\bibfnamefont {G.}~\bibnamefont
  {Lefebvre}}, \bibinfo {author} {\bibfnamefont {T.}~\bibnamefont
  {Antonakakis}}, \bibinfo {author} {\bibfnamefont {Y.}~\bibnamefont
  {Achaoui}}, \bibinfo {author} {\bibfnamefont {R.}~\bibnamefont {Craster}},
  \bibinfo {author} {\bibfnamefont {S.}~\bibnamefont {Guenneau}}, \ and\
  \bibinfo {author} {\bibfnamefont {P.}~\bibnamefont {Sebbah}},\ }\href@noop {}
  {\bibfield  {journal} {\bibinfo  {journal} {Phys. Rev. Lett.}\ }\textbf
  {\bibinfo {volume} {118}},\ \bibinfo {pages} {254302} (\bibinfo {year}
  {2017})}\BibitemShut {NoStop}%
\bibitem [{\citenamefont {Torrent}\ \emph {et~al.}(2013)\citenamefont
  {Torrent}, \citenamefont {Mayou},\ and\ \citenamefont
  {Sanchez-Dehesa}}]{torrent13a}%
  \BibitemOpen
  \bibfield  {author} {\bibinfo {author} {\bibfnamefont {D.}~\bibnamefont
  {Torrent}}, \bibinfo {author} {\bibfnamefont {D.}~\bibnamefont {Mayou}}, \
  and\ \bibinfo {author} {\bibfnamefont {J.}~\bibnamefont {Sanchez-Dehesa}},\
  }\href@noop {} {\bibfield  {journal} {\bibinfo  {journal} {Phys. Rev. B}\
  }\textbf {\bibinfo {volume} {87}},\ \bibinfo {pages} {115143} (\bibinfo
  {year} {2013})}\BibitemShut {NoStop}%
\bibitem [{\citenamefont {Farhat}\ \emph {et~al.}(2009)\citenamefont {Farhat},
  \citenamefont {Guenneau},\ and\ \citenamefont {Enoch}}]{farhat09a}%
  \BibitemOpen
  \bibfield  {author} {\bibinfo {author} {\bibfnamefont {M.}~\bibnamefont
  {Farhat}}, \bibinfo {author} {\bibfnamefont {S.}~\bibnamefont {Guenneau}}, \
  and\ \bibinfo {author} {\bibfnamefont {S.}~\bibnamefont {Enoch}},\
  }\href@noop {} {\bibfield  {journal} {\bibinfo  {journal} {Phys. Rev. Lett.}\
  }\textbf {\bibinfo {volume} {103}},\ \bibinfo {pages} {024301} (\bibinfo
  {year} {2009})}\BibitemShut {NoStop}%
\bibitem [{\citenamefont {Farhat}\ \emph {et~al.}(2010)\citenamefont {Farhat},
  \citenamefont {Guenneau}, \citenamefont {Enoch}, \citenamefont {Movchan},\
  and\ \citenamefont {Petursson}}]{farhat10a}%
  \BibitemOpen
  \bibfield  {author} {\bibinfo {author} {\bibfnamefont {M.}~\bibnamefont
  {Farhat}}, \bibinfo {author} {\bibfnamefont {S.}~\bibnamefont {Guenneau}},
  \bibinfo {author} {\bibfnamefont {S.}~\bibnamefont {Enoch}}, \bibinfo
  {author} {\bibfnamefont {A.}~\bibnamefont {Movchan}}, \ and\ \bibinfo
  {author} {\bibfnamefont {G.}~\bibnamefont {Petursson}},\ }\href@noop {}
  {\bibfield  {journal} {\bibinfo  {journal} {Appl. Phys. Lett.}\ }\textbf
  {\bibinfo {volume} {96}},\ \bibinfo {pages} {081909} (\bibinfo {year}
  {2010})}\BibitemShut {NoStop}%
\bibitem [{\citenamefont {Pal}\ and\ \citenamefont {Ruzzene}(2017)}]{pal17a}%
  \BibitemOpen
  \bibfield  {author} {\bibinfo {author} {\bibfnamefont {R.~K.}\ \bibnamefont
  {Pal}}\ and\ \bibinfo {author} {\bibfnamefont {M.}~\bibnamefont {Ruzzene}},\
  }\href {http://stacks.iop.org/1367-2630/19/i=2/a=025001} {\bibfield
  {journal} {\bibinfo  {journal} {New J. Phys.}\ }\textbf {\bibinfo {volume}
  {19}},\ \bibinfo {pages} {025001} (\bibinfo {year} {2017})}\BibitemShut
  {NoStop}%
\bibitem [{\citenamefont {Rose}\ and\ \citenamefont {Wang}(2004)}]{rose04a}%
  \BibitemOpen
  \bibfield  {author} {\bibinfo {author} {\bibfnamefont {L.~R.~F.}\
  \bibnamefont {Rose}}\ and\ \bibinfo {author} {\bibfnamefont {C.~H.}\
  \bibnamefont {Wang}},\ }\href {\doibase 10.1121/1.1739482} {\bibfield
  {journal} {\bibinfo  {journal} {J. Acoust. Soc. Am.}\ }\textbf {\bibinfo
  {volume} {116}},\ \bibinfo {pages} {154} (\bibinfo {year}
  {2004})}\BibitemShut {NoStop}%
\bibitem [{\citenamefont {Brule}\ \emph {et~al.}(2014)\citenamefont {Brule},
  \citenamefont {Javelaud}, \citenamefont {Enoch},\ and\ \citenamefont
  {Guenneau}}]{brule14}%
  \BibitemOpen
  \bibfield  {author} {\bibinfo {author} {\bibfnamefont {S.}~\bibnamefont
  {Brule}}, \bibinfo {author} {\bibfnamefont {E.}~\bibnamefont {Javelaud}},
  \bibinfo {author} {\bibfnamefont {S.}~\bibnamefont {Enoch}}, \ and\ \bibinfo
  {author} {\bibfnamefont {S.}~\bibnamefont {Guenneau}},\ }\href@noop {}
  {\bibfield  {journal} {\bibinfo  {journal} {Phys. Rev. Lett.}\ }\textbf
  {\bibinfo {volume} {112}} (\bibinfo {year} {2014})}\BibitemShut {NoStop}%
\bibitem [{\citenamefont {Colombi}\ \emph
  {et~al.}(2016{\natexlab{a}})\citenamefont {Colombi}, \citenamefont {Roux},
  \citenamefont {Guenneau}, \citenamefont {Gueguen},\ and\ \citenamefont
  {Craster}}]{colombi16a}%
  \BibitemOpen
  \bibfield  {author} {\bibinfo {author} {\bibfnamefont {A.}~\bibnamefont
  {Colombi}}, \bibinfo {author} {\bibfnamefont {P.}~\bibnamefont {Roux}},
  \bibinfo {author} {\bibfnamefont {S.}~\bibnamefont {Guenneau}}, \bibinfo
  {author} {\bibfnamefont {P.}~\bibnamefont {Gueguen}}, \ and\ \bibinfo
  {author} {\bibfnamefont {R.~V.}\ \bibnamefont {Craster}},\ }\href@noop {}
  {\bibfield  {journal} {\bibinfo  {journal} {Sci. Reps.}\ }\textbf {\bibinfo
  {volume} {6}},\ \bibinfo {pages} {19238} (\bibinfo {year}
  {2016}{\natexlab{a}})}\BibitemShut {NoStop}%
\bibitem [{\citenamefont {Colombi}\ \emph {et~al.}(2014)\citenamefont
  {Colombi}, \citenamefont {Roux},\ and\ \citenamefont {Rupin}}]{colombi14a}%
  \BibitemOpen
  \bibfield  {author} {\bibinfo {author} {\bibfnamefont {A.}~\bibnamefont
  {Colombi}}, \bibinfo {author} {\bibfnamefont {P.}~\bibnamefont {Roux}}, \
  and\ \bibinfo {author} {\bibfnamefont {M.}~\bibnamefont {Rupin}},\
  }\href@noop {} {\bibfield  {journal} {\bibinfo  {journal} {J. Acoust. Soc.
  Am.}\ }\textbf {\bibinfo {volume} {136}},\ \bibinfo {pages} {EL192} (\bibinfo
  {year} {2014})}\BibitemShut {NoStop}%
\bibitem [{\citenamefont {Williams}\ \emph {et~al.}(2015)\citenamefont
  {Williams}, \citenamefont {Roux}, \citenamefont {Rupin},\ and\ \citenamefont
  {Kuperman}}]{williams15a}%
  \BibitemOpen
  \bibfield  {author} {\bibinfo {author} {\bibfnamefont {E.~G.}\ \bibnamefont
  {Williams}}, \bibinfo {author} {\bibfnamefont {P.}~\bibnamefont {Roux}},
  \bibinfo {author} {\bibfnamefont {M.}~\bibnamefont {Rupin}}, \ and\ \bibinfo
  {author} {\bibfnamefont {W.~A.}\ \bibnamefont {Kuperman}},\ }\href@noop {}
  {\bibfield  {journal} {\bibinfo  {journal} {Phys. Rev. B}\ }\textbf {\bibinfo
  {volume} {91}},\ \bibinfo {pages} {104307} (\bibinfo {year}
  {2015})}\BibitemShut {NoStop}%
\bibitem [{\citenamefont {Colombi}\ \emph
  {et~al.}(2016{\natexlab{b}})\citenamefont {Colombi}, \citenamefont
  {Colquitt}, \citenamefont {Roux}, \citenamefont {Guenneau},\ and\
  \citenamefont {Craster}}]{colombi16b}%
  \BibitemOpen
  \bibfield  {author} {\bibinfo {author} {\bibfnamefont {A.}~\bibnamefont
  {Colombi}}, \bibinfo {author} {\bibfnamefont {D.}~\bibnamefont {Colquitt}},
  \bibinfo {author} {\bibfnamefont {P.}~\bibnamefont {Roux}}, \bibinfo {author}
  {\bibfnamefont {S.}~\bibnamefont {Guenneau}}, \ and\ \bibinfo {author}
  {\bibfnamefont {R.~V.}\ \bibnamefont {Craster}},\ }\href@noop {} {\bibfield
  {journal} {\bibinfo  {journal} {Sci. Rep.}\ }\textbf {\bibinfo {volume}
  {6}},\ \bibinfo {pages} {27717} (\bibinfo {year}
  {2016}{\natexlab{b}})}\BibitemShut {NoStop}%
\bibitem [{\citenamefont {Movchan}\ \emph {et~al.}(2007)\citenamefont
  {Movchan}, \citenamefont {Movchan},\ and\ \citenamefont
  {{McPh}edran}}]{movchan07c}%
  \BibitemOpen
  \bibfield  {author} {\bibinfo {author} {\bibfnamefont {A.~B.}\ \bibnamefont
  {Movchan}}, \bibinfo {author} {\bibfnamefont {N.~V.}\ \bibnamefont
  {Movchan}}, \ and\ \bibinfo {author} {\bibfnamefont {R.~C.}\ \bibnamefont
  {{McPh}edran}},\ }\href@noop {} {\bibfield  {journal} {\bibinfo  {journal}
  {Proc. R. Soc. Lond. {\rm A}}\ }\textbf {\bibinfo {volume} {463}},\ \bibinfo
  {pages} {2505} (\bibinfo {year} {2007})}\BibitemShut {NoStop}%
\bibitem [{\citenamefont {Krodel}\ \emph {et~al.}(2015)\citenamefont {Krodel},
  \citenamefont {Thome},\ and\ \citenamefont {Daraio}}]{krodel15a}%
  \BibitemOpen
  \bibfield  {author} {\bibinfo {author} {\bibfnamefont {S.}~\bibnamefont
  {Krodel}}, \bibinfo {author} {\bibfnamefont {N.}~\bibnamefont {Thome}}, \
  and\ \bibinfo {author} {\bibfnamefont {C.}~\bibnamefont {Daraio}},\
  }\href@noop {} {\bibfield  {journal} {\bibinfo  {journal} {Ext. Mech. Lett.}\
  }\textbf {\bibinfo {volume} {4}},\ \bibinfo {pages} {111} (\bibinfo {year}
  {2015})}\BibitemShut {NoStop}%
\bibitem [{\citenamefont {Miniaci}\ \emph {et~al.}(2016)\citenamefont
  {Miniaci}, \citenamefont {Krushynska}, \citenamefont {Bosia},\ and\
  \citenamefont {Pugno}}]{miniaci16a}%
  \BibitemOpen
  \bibfield  {author} {\bibinfo {author} {\bibfnamefont {M.}~\bibnamefont
  {Miniaci}}, \bibinfo {author} {\bibfnamefont {A.}~\bibnamefont {Krushynska}},
  \bibinfo {author} {\bibfnamefont {F.}~\bibnamefont {Bosia}}, \ and\ \bibinfo
  {author} {\bibfnamefont {N.~M.}\ \bibnamefont {Pugno}},\ }\href@noop {}
  {\bibfield  {journal} {\bibinfo  {journal} {New J. Phys.}\ }\textbf {\bibinfo
  {volume} {18}},\ \bibinfo {pages} {083041} (\bibinfo {year}
  {2016})}\BibitemShut {NoStop}%
\bibitem [{\citenamefont {Achaoui}\ \emph {et~al.}(2017)\citenamefont
  {Achaoui}, \citenamefont {Antonakakis}, \citenamefont {Br\^ul\'e},
  \citenamefont {Craster}, \citenamefont {Enoch},\ and\ \citenamefont
  {Guenneau}}]{achaoui17a}%
  \BibitemOpen
  \bibfield  {author} {\bibinfo {author} {\bibfnamefont {Y.}~\bibnamefont
  {Achaoui}}, \bibinfo {author} {\bibfnamefont {T.}~\bibnamefont
  {Antonakakis}}, \bibinfo {author} {\bibfnamefont {S.}~\bibnamefont
  {Br\^ul\'e}}, \bibinfo {author} {\bibfnamefont {R.~V.}\ \bibnamefont
  {Craster}}, \bibinfo {author} {\bibfnamefont {S.}~\bibnamefont {Enoch}}, \
  and\ \bibinfo {author} {\bibfnamefont {S.}~\bibnamefont {Guenneau}},\ }\href
  {http://stacks.iop.org/1367-2630/19/i=6/a=063022} {\bibfield  {journal}
  {\bibinfo  {journal} {New J. Phys.}\ }\textbf {\bibinfo {volume} {19}},\
  \bibinfo {pages} {063022} (\bibinfo {year} {2017})}\BibitemShut {NoStop}%
\bibitem [{\citenamefont {Evans}\ and\ \citenamefont
  {Porter}(2007)}]{evans07a}%
  \BibitemOpen
  \bibfield  {author} {\bibinfo {author} {\bibfnamefont {D.~V.}\ \bibnamefont
  {Evans}}\ and\ \bibinfo {author} {\bibfnamefont {R.}~\bibnamefont {Porter}},\
  }\href@noop {} {\bibfield  {journal} {\bibinfo  {journal} {J. Engng. Math.}\
  }\textbf {\bibinfo {volume} {58}},\ \bibinfo {pages} {317} (\bibinfo {year}
  {2007})}\BibitemShut {NoStop}%
\bibitem [{\citenamefont {Dong}\ \emph {et~al.}(2017)\citenamefont {Dong},
  \citenamefont {Chen}, \citenamefont {Zhu}, \citenamefont {Wang},\ and\
  \citenamefont {Zhang}}]{dong_valley_2017}%
  \BibitemOpen
  \bibfield  {author} {\bibinfo {author} {\bibfnamefont {J.-W.}\ \bibnamefont
  {Dong}}, \bibinfo {author} {\bibfnamefont {X.-D.}\ \bibnamefont {Chen}},
  \bibinfo {author} {\bibfnamefont {H.}~\bibnamefont {Zhu}}, \bibinfo {author}
  {\bibfnamefont {Y.}~\bibnamefont {Wang}}, \ and\ \bibinfo {author}
  {\bibfnamefont {X.}~\bibnamefont {Zhang}},\ }\href {\doibase
  10.1038/nmat4807} {\bibfield  {journal} {\bibinfo  {journal} {Nature
  Materials}\ }\textbf {\bibinfo {volume} {16}},\ \bibinfo {pages} {298}
  (\bibinfo {year} {2017})}\BibitemShut {NoStop}%
\bibitem [{\citenamefont {Zhu}\ \emph {et~al.}(2018)\citenamefont {Zhu},
  \citenamefont {Liu},\ and\ \citenamefont {Semperlotti}}]{zhu_design_2018}%
  \BibitemOpen
  \bibfield  {author} {\bibinfo {author} {\bibfnamefont {H.}~\bibnamefont
  {Zhu}}, \bibinfo {author} {\bibfnamefont {T.-W.}\ \bibnamefont {Liu}}, \ and\
  \bibinfo {author} {\bibfnamefont {F.}~\bibnamefont {Semperlotti}},\
  }\href@noop {} {\bibfield  {journal} {\bibinfo  {journal} {Phys. Rev. B}\
  }\textbf {\bibinfo {volume} {97}} (\bibinfo {year} {2018})},\ \bibinfo {note}
  {arXiv: 1712.10271}\BibitemShut {NoStop}%
\bibitem [{\citenamefont {Xiao}\ \emph {et~al.}(2007)\citenamefont {Xiao},
  \citenamefont {Yao},\ and\ \citenamefont
  {Niu}}]{xiao_valley-contrasting_2007}%
  \BibitemOpen
  \bibfield  {author} {\bibinfo {author} {\bibfnamefont {D.}~\bibnamefont
  {Xiao}}, \bibinfo {author} {\bibfnamefont {W.}~\bibnamefont {Yao}}, \ and\
  \bibinfo {author} {\bibfnamefont {Q.}~\bibnamefont {Niu}},\ }\href
  {https://link.aps.org/doi/10.1103/PhysRevLett.99.236809} {\bibfield
  {journal} {\bibinfo  {journal} {Phys. Rev. Lett.}\ }\textbf {\bibinfo
  {volume} {99}},\ \bibinfo {pages} {236809} (\bibinfo {year}
  {2007})}\BibitemShut {NoStop}%
\bibitem [{\citenamefont {Chen}\ \emph {et~al.}(2017)\citenamefont {Chen},
  \citenamefont {Zhao}, \citenamefont {Chen},\ and\ \citenamefont
  {Dong}}]{chen17a}%
  \BibitemOpen
  \bibfield  {author} {\bibinfo {author} {\bibfnamefont {X.-D.}\ \bibnamefont
  {Chen}}, \bibinfo {author} {\bibfnamefont {F.-L.}\ \bibnamefont {Zhao}},
  \bibinfo {author} {\bibfnamefont {M.}~\bibnamefont {Chen}}, \ and\ \bibinfo
  {author} {\bibfnamefont {J.-W.}\ \bibnamefont {Dong}},\ }\href@noop {}
  {\bibfield  {journal} {\bibinfo  {journal} {Phys. Rev. B}\ }\textbf {\bibinfo
  {volume} {96}},\ \bibinfo {pages} {020202(R)} (\bibinfo {year}
  {2017})}\BibitemShut {NoStop}%
\bibitem [{\citenamefont {Mekis}\ \emph {et~al.}(1996)\citenamefont {Mekis},
  \citenamefont {Chen}, \citenamefont {Kurland}, \citenamefont {Fan},
  \citenamefont {Villeneuve},\ and\ \citenamefont {Joannopoulos}}]{mekis96a}%
  \BibitemOpen
  \bibfield  {author} {\bibinfo {author} {\bibfnamefont {A.}~\bibnamefont
  {Mekis}}, \bibinfo {author} {\bibfnamefont {J.~C.}\ \bibnamefont {Chen}},
  \bibinfo {author} {\bibfnamefont {I.}~\bibnamefont {Kurland}}, \bibinfo
  {author} {\bibfnamefont {S.}~\bibnamefont {Fan}}, \bibinfo {author}
  {\bibfnamefont {P.~R.}\ \bibnamefont {Villeneuve}}, \ and\ \bibinfo {author}
  {\bibfnamefont {J.~D.}\ \bibnamefont {Joannopoulos}},\ }\href@noop {}
  {\bibfield  {journal} {\bibinfo  {journal} {Phys. Rev. Lett.}\ }\textbf
  {\bibinfo {volume} {77}},\ \bibinfo {pages} {3787} (\bibinfo {year}
  {1996})}\BibitemShut {NoStop}%
\bibitem [{\citenamefont {Ma}\ \emph {et~al.}(2015)\citenamefont {Ma},
  \citenamefont {Khanikaev}, \citenamefont {Mousavi},\ and\ \citenamefont
  {Shvets}}]{ma_guiding_2015}%
  \BibitemOpen
  \bibfield  {author} {\bibinfo {author} {\bibfnamefont {T.}~\bibnamefont
  {Ma}}, \bibinfo {author} {\bibfnamefont {A.~B.}\ \bibnamefont {Khanikaev}},
  \bibinfo {author} {\bibfnamefont {S.~H.}\ \bibnamefont {Mousavi}}, \ and\
  \bibinfo {author} {\bibfnamefont {G.}~\bibnamefont {Shvets}},\ }\href
  {https://link.aps.org/doi/10.1103/PhysRevLett.114.127401} {\bibfield
  {journal} {\bibinfo  {journal} {Phys. Rev. Lett.}\ }\textbf {\bibinfo
  {volume} {114}} (\bibinfo {year} {2015})}\BibitemShut {NoStop}%
\bibitem [{\citenamefont {Chutinan}\ \emph {et~al.}(2002)\citenamefont
  {Chutinan}, \citenamefont {Okano},\ and\ \citenamefont
  {Noda}}]{chutinan_wider_2002}%
  \BibitemOpen
  \bibfield  {author} {\bibinfo {author} {\bibfnamefont {A.}~\bibnamefont
  {Chutinan}}, \bibinfo {author} {\bibfnamefont {M.}~\bibnamefont {Okano}}, \
  and\ \bibinfo {author} {\bibfnamefont {S.}~\bibnamefont {Noda}},\ }\href
  {\doibase 10.1063/1.1458529} {\bibfield  {journal} {\bibinfo  {journal}
  {Appl. Phys. Lett.}\ }\textbf {\bibinfo {volume} {80}},\ \bibinfo {pages}
  {1698} (\bibinfo {year} {2002})}\BibitemShut {NoStop}%
\bibitem [{\citenamefont {Xiao}\ \emph {et~al.}(2012)\citenamefont {Xiao},
  \citenamefont {Wen},\ and\ \citenamefont {Wen}}]{xiao12a}%
  \BibitemOpen
  \bibfield  {author} {\bibinfo {author} {\bibfnamefont {Y.}~\bibnamefont
  {Xiao}}, \bibinfo {author} {\bibfnamefont {J.}~\bibnamefont {Wen}}, \ and\
  \bibinfo {author} {\bibfnamefont {X.}~\bibnamefont {Wen}},\ }\href@noop {}
  {\bibfield  {journal} {\bibinfo  {journal} {J. Phys. D: Appl. Phys.}\
  }\textbf {\bibinfo {volume} {45}},\ \bibinfo {pages} {195401} (\bibinfo
  {year} {2012})}\BibitemShut {NoStop}%
\bibitem [{\citenamefont {Johnson}\ and\ \citenamefont
  {Joannopoulos}(2001)}]{johnson01a}%
  \BibitemOpen
  \bibfield  {author} {\bibinfo {author} {\bibfnamefont {S.~G.}\ \bibnamefont
  {Johnson}}\ and\ \bibinfo {author} {\bibfnamefont {J.~D.}\ \bibnamefont
  {Joannopoulos}},\ }\href@noop {} {\bibfield  {journal} {\bibinfo  {journal}
  {Optics Express}\ }\textbf {\bibinfo {volume} {8}},\ \bibinfo {pages} {173}
  (\bibinfo {year} {2001})}\BibitemShut {NoStop}%
\bibitem [{\citenamefont {Jun~Mei}\ and\ \citenamefont
  {Zhang}(2012)}]{Mei_2012}%
  \BibitemOpen
  \bibfield  {author} {\bibinfo {author} {\bibfnamefont {C.~T.~C.}\
  \bibnamefont {Jun~Mei}, \bibfnamefont {Ying~Wu}}\ and\ \bibinfo {author}
  {\bibfnamefont {Z.-Q.}\ \bibnamefont {Zhang}},\ }\href@noop {} {\bibfield
  {journal} {\bibinfo  {journal} {Phys. Rev. B}\ } (\bibinfo {year}
  {2012})}\BibitemShut {NoStop}%
\bibitem [{\citenamefont {Janssen}\ \emph {et~al.}(2016)\citenamefont
  {Janssen}, \citenamefont {Gillet}, \citenamefont {Poncé}, \citenamefont
  {Martin}, \citenamefont {Torrent},\ and\ \citenamefont
  {Gonze}}]{janssen_precise_2016}%
  \BibitemOpen
  \bibfield  {author} {\bibinfo {author} {\bibfnamefont {J.~L.}\ \bibnamefont
  {Janssen}}, \bibinfo {author} {\bibfnamefont {Y.}~\bibnamefont {Gillet}},
  \bibinfo {author} {\bibfnamefont {S.}~\bibnamefont {Poncé}}, \bibinfo
  {author} {\bibfnamefont {A.}~\bibnamefont {Martin}}, \bibinfo {author}
  {\bibfnamefont {M.}~\bibnamefont {Torrent}}, \ and\ \bibinfo {author}
  {\bibfnamefont {X.}~\bibnamefont {Gonze}},\ }\href@noop {} {\bibfield
  {journal} {\bibinfo  {journal} {Phys. Rev. B}\ }\textbf {\bibinfo {volume}
  {93}} (\bibinfo {year} {2016})},\ \bibinfo {note} {arXiv:
  1708.05890}\BibitemShut {NoStop}%
\bibitem [{\citenamefont {Ochiai}(2012)}]{ochiai_photonic_2012}%
  \BibitemOpen
  \bibfield  {author} {\bibinfo {author} {\bibfnamefont {T.}~\bibnamefont
  {Ochiai}},\ }\href {https://link.aps.org/doi/10.1103/PhysRevB.86.075152}
  {\bibfield  {journal} {\bibinfo  {journal} {Phys. Rev. B}\ }\textbf {\bibinfo
  {volume} {86}} (\bibinfo {year} {2012})}\BibitemShut {NoStop}%
\bibitem [{\citenamefont {Lu}\ \emph {et~al.}(2014{\natexlab{c}})\citenamefont
  {Lu}, \citenamefont {Qiu}, \citenamefont {Xu}, \citenamefont {Ye},
  \citenamefont {Ke},\ and\ \citenamefont {Liu}}]{lu_dirac_2014}%
  \BibitemOpen
  \bibfield  {author} {\bibinfo {author} {\bibfnamefont {J.}~\bibnamefont
  {Lu}}, \bibinfo {author} {\bibfnamefont {C.}~\bibnamefont {Qiu}}, \bibinfo
  {author} {\bibfnamefont {S.}~\bibnamefont {Xu}}, \bibinfo {author}
  {\bibfnamefont {Y.}~\bibnamefont {Ye}}, \bibinfo {author} {\bibfnamefont
  {M.}~\bibnamefont {Ke}}, \ and\ \bibinfo {author} {\bibfnamefont
  {Z.}~\bibnamefont {Liu}},\ }\href
  {https://link.aps.org/doi/10.1103/PhysRevB.89.134302} {\bibfield  {journal}
  {\bibinfo  {journal} {Phys. Rev. B}\ }\textbf {\bibinfo {volume} {89}}
  (\bibinfo {year} {2014}{\natexlab{c}})}\BibitemShut {NoStop}%
\bibitem [{\citenamefont {Zhang}\ and\ \citenamefont
  {Niu}(2017)}]{Zhang_Niu_2017}%
  \BibitemOpen
  \bibfield  {author} {\bibinfo {author} {\bibfnamefont {L.}~\bibnamefont
  {Zhang}}\ and\ \bibinfo {author} {\bibfnamefont {Q.}~\bibnamefont {Niu}},\
  }\href@noop {} {\bibfield  {journal} {\bibinfo  {journal} {Phys. Rev. Lett.}\
  } (\bibinfo {year} {2017})}\BibitemShut {NoStop}%
\bibitem [{\citenamefont {Liu}\ \emph {et~al.}(2017)\citenamefont {Liu},
  \citenamefont {Xu}, \citenamefont {Zhang},\ and\ \citenamefont
  {Duan}}]{Liu_Xu_Zhang_2017}%
  \BibitemOpen
  \bibfield  {author} {\bibinfo {author} {\bibfnamefont {Y.}~\bibnamefont
  {Liu}}, \bibinfo {author} {\bibfnamefont {Y.}~\bibnamefont {Xu}}, \bibinfo
  {author} {\bibfnamefont {S.-C.}\ \bibnamefont {Zhang}}, \ and\ \bibinfo
  {author} {\bibfnamefont {W.}~\bibnamefont {Duan}},\ }\href@noop {} {\bibfield
   {journal} {\bibinfo  {journal} {Phys. Rev. B}\ }\textbf {\bibinfo {volume}
  {96}},\ \bibinfo {pages} {064106} (\bibinfo {year} {2017})}\BibitemShut
  {NoStop}%
\bibitem [{\citenamefont {Liu}\ \emph {et~al.}(2018)\citenamefont {Liu},
  \citenamefont {Xu},\ and\ \citenamefont {Duan}}]{Liu_Xu_2017}%
  \BibitemOpen
  \bibfield  {author} {\bibinfo {author} {\bibfnamefont {Y.}~\bibnamefont
  {Liu}}, \bibinfo {author} {\bibfnamefont {Y.}~\bibnamefont {Xu}}, \ and\
  \bibinfo {author} {\bibfnamefont {W.}~\bibnamefont {Duan}},\ }\href@noop {}
  {\bibfield  {journal} {\bibinfo  {journal} {National Science Review}\
  }\textbf {\bibinfo {volume} {5}},\ \bibinfo {pages} {314–} (\bibinfo {year}
  {2018})}\BibitemShut {NoStop}%
\bibitem [{\citenamefont {Guo}\ and\ \citenamefont {Franz}(2009)}]{Guo_2009}%
  \BibitemOpen
  \bibfield  {author} {\bibinfo {author} {\bibfnamefont {H.}~\bibnamefont
  {Guo}}\ and\ \bibinfo {author} {\bibfnamefont {M.}~\bibnamefont {Franz}},\
  }\href@noop {} {\bibfield  {journal} {\bibinfo  {journal} {Phys. Rev. B}\
  }\textbf {\bibinfo {volume} {80}},\ \bibinfo {pages} {113102} (\bibinfo
  {year} {2009})}\BibitemShut {NoStop}%
\bibitem [{\citenamefont {Ni}\ \emph {et~al.}(2017)\citenamefont {Ni},
  \citenamefont {Gorlach}, \citenamefont {Alu},\ and\ \citenamefont
  {Khanikaev}}]{ni_topological_2017}%
  \BibitemOpen
  \bibfield  {author} {\bibinfo {author} {\bibfnamefont {X.}~\bibnamefont
  {Ni}}, \bibinfo {author} {\bibfnamefont {M.~A.}\ \bibnamefont {Gorlach}},
  \bibinfo {author} {\bibfnamefont {A.}~\bibnamefont {Alu}}, \ and\ \bibinfo
  {author} {\bibfnamefont {A.~B.}\ \bibnamefont {Khanikaev}},\ }\href {\doibase
  10.1088/1367-2630/aa6996} {\bibfield  {journal} {\bibinfo  {journal} {New J.
  Phys}\ }\textbf {\bibinfo {volume} {19}},\ \bibinfo {pages} {055002}
  (\bibinfo {year} {2017})}\BibitemShut {NoStop}%
\bibitem [{\citenamefont {Mei}\ \emph {et~al.}(2012)\citenamefont {Mei},
  \citenamefont {Wu}, \citenamefont {Chan},\ and\ \citenamefont
  {Zhang}}]{mei_first-principles_2012}%
  \BibitemOpen
  \bibfield  {author} {\bibinfo {author} {\bibfnamefont {J.}~\bibnamefont
  {Mei}}, \bibinfo {author} {\bibfnamefont {Y.}~\bibnamefont {Wu}}, \bibinfo
  {author} {\bibfnamefont {C.~T.}\ \bibnamefont {Chan}}, \ and\ \bibinfo
  {author} {\bibfnamefont {Z.-Q.}\ \bibnamefont {Zhang}},\ }\href@noop {}
  {\bibfield  {journal} {\bibinfo  {journal} {Phys. Rev. B}\ }\textbf {\bibinfo
  {volume} {86}},\ \bibinfo {pages} {035141} (\bibinfo {year}
  {2012})}\BibitemShut {NoStop}%
\bibitem [{\citenamefont {Berry}(1984)}]{Berry_1983}%
  \BibitemOpen
  \bibfield  {author} {\bibinfo {author} {\bibfnamefont {M.}~\bibnamefont
  {Berry}},\ }\href@noop {} {\bibfield  {journal} {\bibinfo  {journal} {Proc.
  R. Soc. Lond. A}\ }\textbf {\bibinfo {volume} {392}},\ \bibinfo {pages} {45}
  (\bibinfo {year} {1984})}\BibitemShut {NoStop}%
\bibitem [{\citenamefont {Xiao}\ \emph {et~al.}(2010)\citenamefont {Xiao},
  \citenamefont {Chang},\ and\ \citenamefont {Niu}}]{Xiao_2010}%
  \BibitemOpen
  \bibfield  {author} {\bibinfo {author} {\bibfnamefont {D.}~\bibnamefont
  {Xiao}}, \bibinfo {author} {\bibfnamefont {M.-C.}\ \bibnamefont {Chang}}, \
  and\ \bibinfo {author} {\bibfnamefont {Q.}~\bibnamefont {Niu}},\ }\href@noop
  {} {\bibfield  {journal} {\bibinfo  {journal} {Rev. Mod. Phys.}\ }\textbf
  {\bibinfo {volume} {82}},\ \bibinfo {pages} {1959} (\bibinfo {year}
  {2010})}\BibitemShut {NoStop}%
\bibitem [{\citenamefont {Vasseur}\ \emph {et~al.}(2013)\citenamefont
  {Vasseur}, \citenamefont {{Fagot-Revurat}}, \citenamefont {Kierren},
  \citenamefont {Sicot},\ and\ \citenamefont {Malterre}}]{vasseur13a}%
  \BibitemOpen
  \bibfield  {author} {\bibinfo {author} {\bibfnamefont {G.}~\bibnamefont
  {Vasseur}}, \bibinfo {author} {\bibfnamefont {Y.}~\bibnamefont
  {{Fagot-Revurat}}}, \bibinfo {author} {\bibfnamefont {B.}~\bibnamefont
  {Kierren}}, \bibinfo {author} {\bibfnamefont {M.}~\bibnamefont {Sicot}}, \
  and\ \bibinfo {author} {\bibfnamefont {D.}~\bibnamefont {Malterre}},\
  }\href@noop {} {\bibfield  {journal} {\bibinfo  {journal} {Symmetry}\
  }\textbf {\bibinfo {volume} {5}},\ \bibinfo {pages} {344} (\bibinfo {year}
  {2013})}\BibitemShut {NoStop}%
\bibitem [{\citenamefont {Min}\ \emph {et~al.}(2006)\citenamefont {Min},
  \citenamefont {Wu}, \citenamefont {Zhong}, \citenamefont {Zhong},
  \citenamefont {Zhong},\ and\ \citenamefont {Liu}}]{min06a}%
  \BibitemOpen
  \bibfield  {author} {\bibinfo {author} {\bibfnamefont {R.}~\bibnamefont
  {Min}}, \bibinfo {author} {\bibfnamefont {F.}~\bibnamefont {Wu}}, \bibinfo
  {author} {\bibfnamefont {L.}~\bibnamefont {Zhong}}, \bibinfo {author}
  {\bibfnamefont {H.}~\bibnamefont {Zhong}}, \bibinfo {author} {\bibfnamefont
  {S.}~\bibnamefont {Zhong}}, \ and\ \bibinfo {author} {\bibfnamefont
  {Y.}~\bibnamefont {Liu}},\ }\href
  {http://stacks.iop.org/0022-3727/39/i=10/a=041} {\bibfield  {journal}
  {\bibinfo  {journal} {J. Phys. D: Appl. Phys.}\ }\textbf {\bibinfo {volume}
  {39}},\ \bibinfo {pages} {2272} (\bibinfo {year} {2006})}\BibitemShut
  {NoStop}%
\bibitem [{\citenamefont {Wang}\ \emph {et~al.}(2007)\citenamefont {Wang},
  \citenamefont {Li}, \citenamefont {Huang},\ and\ \citenamefont
  {Wang}}]{wang07a}%
  \BibitemOpen
  \bibfield  {author} {\bibinfo {author} {\bibfnamefont {Y.-Z.}\ \bibnamefont
  {Wang}}, \bibinfo {author} {\bibfnamefont {F.-M.}\ \bibnamefont {Li}},
  \bibinfo {author} {\bibfnamefont {W.-H.}\ \bibnamefont {Huang}}, \ and\
  \bibinfo {author} {\bibfnamefont {Y.-S.}\ \bibnamefont {Wang}},\ }\href@noop
  {} {\bibfield  {journal} {\bibinfo  {journal} {J. Phys: Cond. Mat.}\ }\textbf
  {\bibinfo {volume} {19}},\ \bibinfo {pages} {496204} (\bibinfo {year}
  {2007})}\BibitemShut {NoStop}%
\bibitem [{\citenamefont {Wang}\ \emph {et~al.}(2011)\citenamefont {Wang},
  \citenamefont {Wang},\ and\ \citenamefont {Su}}]{wang11a}%
  \BibitemOpen
  \bibfield  {author} {\bibinfo {author} {\bibfnamefont {Y.-F.}\ \bibnamefont
  {Wang}}, \bibinfo {author} {\bibfnamefont {Y.-S.}\ \bibnamefont {Wang}}, \
  and\ \bibinfo {author} {\bibfnamefont {X.-X.}\ \bibnamefont {Su}},\
  }\href@noop {} {\bibfield  {journal} {\bibinfo  {journal} {J. Appl. Phys.}\
  }\textbf {\bibinfo {volume} {110}},\ \bibinfo {pages} {113520} (\bibinfo
  {year} {2011})}\BibitemShut {NoStop}%
\bibitem [{\citenamefont {Wang}\ \emph {et~al.}(2001)\citenamefont {Wang},
  \citenamefont {Wang}, \citenamefont {Gu},\ and\ \citenamefont
  {Yang}}]{wang01a}%
  \BibitemOpen
  \bibfield  {author} {\bibinfo {author} {\bibfnamefont {R.}~\bibnamefont
  {Wang}}, \bibinfo {author} {\bibfnamefont {X.-H.}\ \bibnamefont {Wang}},
  \bibinfo {author} {\bibfnamefont {B.-Y.}\ \bibnamefont {Gu}}, \ and\ \bibinfo
  {author} {\bibfnamefont {G.-Z.}\ \bibnamefont {Yang}},\ }\href@noop {}
  {\bibfield  {journal} {\bibinfo  {journal} {J. Appl. Phys.} \bibinfo
  {pages} {4307}} (\bibinfo {year} {2001})}\BibitemShut {NoStop}%
\bibitem [{\citenamefont {Jin}\ \emph {et~al.}(2018)\citenamefont {Jin},
  \citenamefont {Torrent}, \ and\ \citenamefont
  {Djafari-Rouhani}}]{jin18a}%
 \BibitemOpen
  \bibfield  {author} {\bibinfo {author} {\bibfnamefont {Y.}~\bibnamefont
  {Jin}}, \bibinfo {author} {\bibfnamefont {D.}\ \bibnamefont {Torrent}},
  \bibinfo {author} {\bibfnamefont {B.}\ and\ \bibnamefont {Djafari-Rouhani}},\ }\href@noop {}
  {\bibfield  {journal} {\bibinfo  {journal} {Phys. Rev. B}\ \bibinfo
  {pages} {054307}} (\bibinfo {year} {2018})}\BibitemShut {NoStop}%
\end{thebibliography}

%

\section{Appendices}

\appendix

\section{Character tables}
\label{sec:char_tables}



\onecolumngrid

\begin{table}[h!]
\centering
\begin{tabular}{|l|ccc|cc|}
\hline
\cellcolor[HTML]{EFEFEF}Classes $\rightarrow$ &                       &                          &                               & \multicolumn{2}{c|}{\hspace{-1.5cm}Basis Functions}                                                                                                                         \\
\cellcolor[HTML]{EFEFEF}IR $\downarrow$       & \multirow{-2}{*}{$E$} & \multirow{-2}{*}{$2C_3$} & \multirow{-2}{*}{$3\sigma_v$} & Odd-Parity                                                                                  & Even-Parity                                                    \\ \hline
                                              &                       &                          &                               & \multicolumn{1}{c|}{$\{x, y \}$}                                                            & \begin{tabular}[c]{@{}c@{}}$\{x^2-y^2, 2xy \}$\end{tabular} \\
\multirow{-2}{*}{$E$}                         & \multirow{-2}{*}{$2$} & \multirow{-2}{*}{$-1$}   & \multirow{-2}{*}{$0$}         & \multicolumn{1}{c|}{\begin{tabular}[c]{@{}c@{}}$\{h_1(x, y), h_2(x, y) \}$\end{tabular}} & $\{(x^2-y^2)^2-4x^2y^2, xy(x^2-y^2) \}$                        \\ \hline
\end{tabular}
\caption{Selected row and basis functions of $C_{3v}$ character table.}
\label{C3v}
\end{table}
\twocolumngrid

\onecolumngrid

\begin{table}[h!]
\centering
\begin{tabular}{|l|ccc|cc|}
\hline
\cellcolor[HTML]{EFEFEF}Classes $\rightarrow$ &                       &                          &                               & \multicolumn{2}{c|}{\hspace{-1.5cm}Basis Functions}                                                                                                                         \\
\cellcolor[HTML]{EFEFEF}IR $\downarrow$       & \multirow{-2}{*}{$E$} & \multirow{-2}{*}{$C_3$} & \multirow{-2}{*}{$C^2_3$} & Odd-Parity                                                                                  & Even-Parity                                                    \\ \hline
                                              & $1$                   & $\epsilon$               & $\epsilon^2$                  & \multicolumn{1}{c|}{$\{x, y \}$}                                                            & \begin{tabular}[c]{@{}c@{}}$\{x^2-y^2, 2xy \}$\end{tabular} \\
\multirow{-2}{*}{$E$}                         & $1$                   & $\epsilon^2$             & $\epsilon$                    & \multicolumn{1}{c|}{\begin{tabular}[c]{@{}c@{}}$\{h_1(x, y), h_2(x, y) \}$\end{tabular}} & $\{(x^2-y^2)^2-4x^2y^2, xy(x^2-y^2) \}$                        \\ \hline
\end{tabular}
\caption{Selected row and basis functions of $C_{3}$ character table; $\epsilon = \exp\left(2\pi i/3 \right)$.}
\label{C3}
\end{table}
\twocolumngrid

\section{Irreducible representation matrices}
\label{sec:IMR}
IMR for three-fold rotations:
\beq
D_E(\pm C_3) = \left[ \begin{array}{rr}
-1/2 & \mp \sqrt{3}/2  \\
\pm\sqrt{3}/2 & -1/2  \end{array} \right].
\label{eq:IR_E_C3_C3V} \eeq This IMR is for the odd-parity basis $\{x, y\}$ which is chosen because it simplifies equation \eqref{eq:matrix_elts_relationship} moreso than its even-parity counterpart $\{x^2 - y^2, 2xy\}$; due to the property $\pm\hat{C_3} = D_E({\pm C_3})$. As we are seeking to examine the local behaviour in the vicinity of the Dirac cone, either parity basis may be used, as they are both associated to the same IR.
\\

The IMR of $\sigma_v$, shown in table \ref{C3v}, is explicitly,
\beq
D_E(\sigma_v) = -\hat{\sigma}_{z},
\label{eq:IR_E_sigmav} \eeq where $\hat{\sigma}_z$ is the third Pauli matrix; this representation implies that $\sigma_v$ is the parity transformation $\kappa_y \rightarrow -\kappa_y$.
\\

We have shown that for $KK^\prime$ valleys with $C_3$ symmetry we get a determistic Dirac cone if and only if $G_{\Gamma} = C_{6}$. In this instance, the sticking together of the two one-dimensional complex representations yields IMR's of the form,
\beq
D'_E(R) = \left[ \begin{array}{rr}
\omega & 0  \\
0 & \omega^2  \end{array} \right],
\eeq which is equivalent to the $C_{3v}$ rotation matrices after undergoing an equivalence transformation,
\beq
D_E(R) = U D'_E(R) U^{-1},\quad U = \frac{1}{\sqrt{2}}\left[ \begin{array}{rr}
i & -i  \\
1 & 1  \end{array} \right],
\eeq therefore the form of the perturbed Hamiltonian, after reduction using rotational symmetries, is identical for systems which have $C_3$ and $C_{3v}$ point group symmetry at $KK^\prime$.

\end{document}